\documentclass{emulateapj}

\usepackage{epsfig}

\shorttitle{MHD HII Region Evolution}
\shortauthors{Krumholz, Stone, \& Gardiner}

\newcommand{\ltsim}{\protect\raisebox{-0.5ex}{$\:\stackrel{\textstyle <}
        {\sim}\:$}}
\newcommand{\gtsim}{\protect\raisebox{-0.5ex}{$\:\stackrel{\textstyle >}
        {\sim}\:$}}

\newcommand{\vecx}{\mathbf{x}}
\newcommand{\vecv}{\mathbf{v}}
\newcommand{\vecB}{\mathbf{B}}
\newcommand{\vecp}{\mathbf{p}}
\newcommand{\vecq}{\mathbf{q}}
\newcommand{\rhon}{\rho_{n}}
\newcommand{\calg}{\mathcal{G}}
\newcommand{\call}{\mathcal{L}}
\newcommand{\calr}{\mathcal{R}}
\newcommand{\cali}{\mathcal{I}}
\newcommand{\alphaB}{\alpha^{(B)}}
\newcommand{\alphaC}{\alpha_C}
\newcommand{\nHplus}{n_{H^+}}
\newcommand{\eGamma}{e_{\Gamma}}
\newcommand{\muH}{\mu_H}

\begin{document}

\title{Magnetohydronamic Evolution of HII Regions in Molecular Clouds:
Simulation Methodology, Tests, and Uniform Media}

\slugcomment{Accepted to the Astrophysical Journal, August 22, 2007}

\author{Mark R. Krumholz\altaffilmark{1}, James M. Stone, and Thomas A. Gardiner}
\affil{Department of Astrophysical Sciences, Princeton University,
Princeton, NJ 08544}

\altaffiltext{1}{Hubble Fellow}

\begin{abstract}
We present a method for simulating the evolution of HII regions driven
by point sources of ionizing radiation in magnetohydrodynamic media,
implemented in the three-dimensional Athena MHD code. We compare
simulations using our algorithm to analytic solutions and show that
the method passes rigorous tests of accuracy and
convergence. The tests reveal several conditions that an
ionizing radiation-hydrodynamic code must satisfy to reproduce
analytic solutions. As a demonstration of our new method, we
present the first three-dimensional, global simulation of an HII region
expanding into a magnetized gas. The simulation shows that magnetic
fields suppress sweeping up of gas perpendicular to magnetic field
lines, leading to small density contrasts and extremely weak shocks at
the leading edge of the HII region's expanding shell.
\end{abstract}


\section{Introduction}

Observations show that giant molecular clouds (GMCs) in local group
galaxies convert at most a few percent of their mass into stars per
cloud crossing time \citep{zuckerman74}, and that clouds are typically
destroyed in a few crossing times \citep{blitz06a}, well before they
have converted a significant fraction of their mass into
stars. Observations also strongly support the idea that GMCs are
gravitationally bound \citep{krumholz05c,blitz06a,krumholz06d}, so the
fact that they survive for more than a crossing time yet do not
collapse entirely into stars strongly suggests that internal feedback
plays a dominant role in GMC evolution. HII regions driven by
newly-formed massive stars are likely to be the dominant sources of
energy injection and mass loss in clouds, and are therefore critical
to GMC evolution
\citep{mckee97,williams97,matzner02}. \citet{krumholz06d} show using
semi-analytic models that feedback from HII regions can quantiatively
reproduce the observed lifetime, star formation rate, and star
formation efficiency of GMCs. However, these calculations rely on
simple analytic solutions for the evolution of spherically-symmetric
HII regions in non-magnetic gas. Understanding the detailed evolution
of GMCs will require a considerably more sophisticated numerical
approach, and for this reason three-dimensional simulation of the
evolution of molecular clouds under the influence of internal sources
of ionizing radiation is a critical problem in numerical astrophysics.

In this paper we have the dual purpose of presenting a new algorithm
for simulation of HII regions in molecular clouds, and using this
algorithm to explore potential computational problems that arise in
general for simulations of this type. We also demonstrate our new
algorithm in a simple application, the expansion of an HII region into
a uniform neutral gas in which the magnetic pressure greatly exceeds
the thermal pressure, as it does in molecular clouds. In the past year
several authors have considered the problem of simulating HII regions, and
presented both algorithms \citep[e.g.][]{arthur06, mellema06} and
results on the evolution of HII regions in turbulent hydrodynamic
media \citep{dale05, mellema05,maclow06}. For a recent review see
\citet{henney06}. Several authors have also presented methods
for simulation of ionizing radiation-hydrodynamics (IRHD) in a
cosmological context \citep[e.g.][]{abel02, whalen06}, which
differs from the problem in the context of present-day molecular
clouds primarily in the amount of cooling to which the gas is
subjected and the conditions of the gas before it is ionized. Finally,
researchers studying the evolution of ultracompact HII regions
planetary nebulae have presented algorithms and results for ionizing
radiative transfer with hydrodynamics and magnetohydrodynamics in 2D
and 3D under the simplifying assumption that the ionized gas has a
perfectly sharp edge and is always in thermal and ionization
equilibrium \citep[e.g.][]{garciasegura96, garciasegura97,
garciasegura00}.

Our algorithm improves on previous work in that it is the first to
couple ionizing radiative transfer to magnetohydrodynamics
(IRMHD) without imposing an assumption of thermal or ionization
equilibrium. The inclusion of magnetic fields is important because
observations indicate that the magnetic energy in molecular clouds is
comparable to the kinetic and gravitational potential energies
\citep{crutcher99,crutcher05,heiles05}. Thus, dynamical expansion of
ionized regions may be significantly altered by magnetic
confinement, an effect we wish to explore. Indeed, we show that even
in the simple case of expansion of an HII region into a uniform,
magnetized medium, the magnetic field produces qualitatively new
phenomena. The Alfven speeds of a few km s$^{-1}$ typically found in
molecular clouds reduce
the strength of shocks associated with expanding ionization fronts at
early times, and at later times turn off the shocks entirely,
greatly reducing the collection of gas and possibly thereby reducing
the amount of triggered star formation. A non-equilibrium treatment of
the ionization structure is important because we find that a
non-equilibrium treatment of the thermal and ionization structure of
the ionized gas leads to modest but significant effects on quantities
such as the expansion rate of HII regions, effects that are lost in
the assumption of perfect radiative equilibrium.

Before exploring these new phenomena, however, we point out that there
has yet to be a detailed study of potential
computational problems and constraints that arise in 3D simulations of
IRHD and IRMHD with the strong cooling and large temperature contrasts
that are expected for modern day (as opposed to primordial)
interstellar chemistry. We have therefore performed a detailed
comparison of simulations using our method to analytic solutions for
computationally challenging problems, as a way of searching for
potential difficulties that may arise in general IRHD methods. We
discover several conditions that a simulation must satisfy to
reproduce analytic results correctly. One must limit the amount by
which the gas pressure is allowed to change between hydrodynamic
updates, for a ray-tracing method one must periodically rotate the
orientation of the rays, and one must either resolve the ionization
front or restrict the rate of cooling at the front to suppress excess
cooling due to numerical mixing. We show that
failure to meet these conditions produces quantitatively incorrect
results.

The remainder of this paper proceeds as follows. In
\S~\ref{formulation} we describe the physical formulation of the
problem that we adopt, including our approximations and
assumptions. In \S~\ref{algorithm} we present our simulation
algorithm. In \S~\ref{tests} we compare our code to analytic
solutions, and thereby demonstrate the existence of conditions that
numerical methods must satisfy in order to reproduce analytic
solutions correctly. We use our method to simulate the
evolution of HII regions in magnetized media in \S~\ref{MHDresults},
and finally we summarize and present conclusions in \S~\ref{conclusion}.

\section{Physical Formulation}
\label{formulation}

We wish to simulate the evolution of a magnetized molecular gas subjected
to an ionizing radiation field due to point sources within it. The
$n$ sources of ionization are located at positions $\vecx_n$, and each
has an ionizing luminosity of $s_n$, in units of photons per
unit time above the Lyman limit. We treat the ionizing sources as
monochromatic, and assume that they emit all their ionizing photons at
frequencies near the ionization threshhold of hydrogen. Since our
focus is on the dynamics of the cloud rather than the detailed
chemistry of the ionized gas, this approximation is quite reasonable.

We describe the gas by a total mass density $\rho$, density of
neutral species $\rhon$, velocity $\vecv$, and total energy density
$E$. The gas is threaded by a magnetic field $\vecB$. The interface
between an HII region and a molecular cloud is a photodissociation
region (PDR), within which there is a significant amount of atomic
gas. In principle we should therefore separately track ionized,
atomic, and molecular hydrogen, and possibly other species as
well. However, PDRs do not have a signicant effect on the
large-scale dynamics of the ionized gas unless the ionizing source is
very weak or the neutral gas is much less dense than as typical for
molecular clouds, as we show in \S~\ref{PDRs}. For this reason, we
neglect atomic gas, and assume that the density of molecular gas is
$\rhon$ and the density of ionized gas is $\rho-\rhon$.

\subsection{Magnetohydrodynamics}

We compute the evolution of our gas using the equations of
multi-species ideal MHD including radiative heating and cooling and
chemical evolution terms. In conservative form, these are
\begin{eqnarray}
\label{massconservation}
\frac{\partial \rho}{\partial t} + \nabla\cdot(\rho\vecv) & = & 0 \\
\label{momentumconservation}
\frac{\partial}{\partial t}(\rho \vecv) + \nabla\cdot
(\rho \vecv \vecv - \vecB \vecB) + \nabla P^* & = & 0 \\
\label{fluxfreezing}
\frac{\partial \vecB}{\partial t} + \nabla \cdot
(\vecv\vecB - \vecB\vecv) & = & 0 \\
\label{energyconservation}
\frac{\partial E}{\partial t} + \nabla \cdot
\left[(E + P^*)\vecv - \vecB(\vecB\cdot\vecv)\right] & = & \calg -
\call \\
\label{neutralconservation}
\frac{\partial \rhon}{\partial t} + \nabla\cdot(\rhon\vecv) & = &
\calr - \cali,
\end{eqnarray}
where $P^*\equiv P + (\vecB\cdot\vecB)/2$ is the total pressure, $P$
is the gas thermal pressure, $\calr$ and $\cali$ are the rates of
recombination and ionization, measured in mass per unit time per
unit volume, and $\calg$ and $\call$ are the rates of radiative
heating and cooling, in energy per unit time per unit volume. The
total energy density $E$ not including atomic or molecular binding
energies is
\begin{equation}
E \equiv \epsilon + \rho \frac{\vecv\cdot\vecv}{2} +
\frac{\vecB\cdot\vecB}{2},
\end{equation}
where $\epsilon$ is the gas thermal energy density. We adopt an ideal
gas equation of state, so the pressure is $P = (\gamma-1) \epsilon$
with $\gamma=5/3$, corresponding to a monatomic gas. This choice of
$\gamma$ is appropriate because ionized gas is monatomic, and
hydrogen in molecular clouds, while diatomic, is generally too cool to
access its rotational or vibrational degrees of freedom. It therefore
acts as if it were monatomic for the purpose of dynamics. In addition to
the evolution equations, the magnetic field is subject to the
divergence-free constraint
\begin{equation}
\label{divbconstraint}
\nabla\cdot\vecB = 0.
\end{equation}
Physically, equations
(\ref{massconservation}), (\ref{momentumconservation}), and 
(\ref{energyconservation}) represent
conservation of mass, momentum, and energy, equation
(\ref{fluxfreezing}) represents the approximation that the magnetic
field is perfectly frozen in to the fluid, and equation
(\ref{neutralconservation}) states that the mass of neutral gas is
conserved by advection, and is altered only by ionizations and
recombinations. Equation
(\ref{divbconstraint}) expresses the non-existence of magnetic
monopoles. Note that we have adopted a system of units in which the
magnetic permeability $\mu=1$; to convert to cgs units, we multiply
our magnetic field strengths by a factor of $\sqrt{4\pi}$.

\subsection{Recombination and Ionization}

To complete the evolution equations, we must specify the rates of
the radiative processes: recombination, ionization, heating, and
cooling. Recombination is the simplest of these. We adopt the
on-the-spot approximation whereby any recombination to the ground
state of hydrogen is assumed to yield an ionizing photon with a very
short mean free path, which will in turn cause another ionization at
essentially the same location as the recombination
\citep{osterbrock89}. On the other hand, recombinations to excited
states of hydrogen produce photons with long mean free paths, which
are assumed to escape the HII region. Thus, the recombination rate is
approximately
\begin{equation}
\label{recomb}
\calr = \muH \alphaB n_e \nHplus,
\end{equation}
where $\muH\approx 2.34\times 10^{-24}$ g is the mean gas mass per
hydrogen atom (for a gas of hydrogen and helium in the standard cosmic
abundance), $n_e$ and $\nHplus$ are
the number densities of electrons and hydrogen nuclei, and $\alphaB$
is the case B recombination coefficient, $\alphaB \approx 2.59\times
10^{-13} (T/10^4\mbox{ K})^{-0.7}$ cm$^{-3}$ s$^{-1}$
\citep{osterbrock89, rijkhorst05} at a gas temperature $T$. The
H$^{+}$ and e$^{-}$ number densities are $\nHplus
= (\rho-\rhon)/\muH$ and $n_e = \nHplus + \rho \alphaC / (14 \muH)$,
where $\alphaC \approx 3\times 10^{-3}$ is the carbon abundance in the
ISM \citep{sofia01}. This assumes that all carbon is singly-ionized.
Since the first ionization potential of carbon is smaller than that of
hydrogen, this is likely to be the case in regions where there is any
ionized hydrogen present, which are the only regions for which we care
about the recombination rate.

In the on-the-spot approximation, the ionization rate is relatively
easy to compute as well, since we need only solve the radiative
transfer equation only along rays from ionizing sources. The ionizing
flux at position $\vecx$ due to the $n$th ionizing source is
\begin{equation}
F_n = \frac{s_n}{4\pi\left|\vecx-\vecx_n\right|^2}
e^{-\tau(\vecx,\vecx_n)},
\end{equation}
where
\begin{equation}
\tau(\vecx,\vecx_n) \approx \int_{\vecx_n}^{\vecx} (\sigma n_H
+ \sigma_d n) \, d\ell
\end{equation}
is the optical depth to ionizing photons along the path from $\vecx_n$
to $\vecx$, $n_H = \rhon / \muH$ is the number density of neutral
hydrogen atoms, $n=\rho/\muH$ is the number density of hydrogen
nuclei, $\sigma = 6.3\times 10^{-18}$ cm$^2$ is the
cross section for absorption of a photon at the ionization threshhold
by a neutral hydrogen atom, and $\sigma_d\approx 6.1 \times 10^{-22}$
cm$^{-2}$ per H atom is the cross section for absorption of ionizing photons by
dust grains \citep{bertoldi96}. For the purposes of the tests we
report in this paper, for which we are comparing to analytic solutions
derived for pure hydrogen, we set $\sigma_d=0$.
The photoionization rate is just the rate
at which these photons are absorbed, so the total photoionization rate
due to all sources at position $x$ is
\begin{equation}
\cali_{\rm ph} =  \sigma \rhon
\sum_n \frac{s_n}{4\pi\left|\vecx-\vecx_n\right|^2}
e^{-\tau(\vecx,\vecx_n)}.
\end{equation}
We also compute the rate of ionizations due to collisions between
neutral hydrogen atoms and electrons,
\begin{equation}
\label{collion}
\cali_{\rm coll} \approx k_{\rm coll} n_e \rhon,
\end{equation}
where the collisional ionization rate coefficient is
\citep{tenoriotagle86}
\begin{equation}
k_{\rm coll} \approx 5.84\times 10^{-11} \sqrt{\frac{T}{\mbox{ K}}}
\,e^{(-13.6\mbox{ \scriptsize eV})/(k_B T)}\mbox{ cm}^3\mbox{ s}^{-1},
\end{equation}
although for the gas densities and temperatures in the systems with
which we are concerned here this term is generally negligible.
The total ionization rate is the sum of these two terms, 
\begin{equation}
\cali = \rhon \left(\sigma
\sum_n \frac{s_n}{4\pi\left|\vecx-\vecx_n\right|^2}
e^{-\tau(\vecx,\vecx_n)} + k_{\rm coll} n_e\right).
\end{equation}

It should be noted that \citet{ritzerveld05} has recently pointed out
that diffuse photons (i.e. those that have been absorbed and
re-emitted) can be as important as direct photons in
determining the structure of HII regions, so the on-the-spot
approximation we adopt is potentially problematic. However, the
importance of diffuse versus direct photons depends strongly on the
degree of central concentration of the gas inside the HII region. Once
an HII region becomes D type, which is the phase in which we are most
interested from the standpoint of radiation hydrodynamic evolution,
the gas inside the ionized region is nearly uniform in density. For
this case, \citet{ritzerveld05} finds that direct photons dominate
diffuse ones over the majority of the HII region volume, and the
on-the-spot approximation is reasonable. Nonetheless, the on-the-spot
approximation does make the shadows cast by dense gas sharper than
they should be in reality, and we must keep this limitation in mind.

\subsection{Heating and Cooling}

We can approximately determine the heating and cooling rates by
breaking them up into heating and cooling associated with ionization
and recombination of hydrogen atoms, and all other sources of heating
and cooling. The heating rates due to ionization and recombination of
hydrogen are quite simple in our monochromatic treatment of the
ionizing radiation. In this approximation, each absorption of an
ionizing photon delivers $\eGamma \approx 2.4$ eV of thermal energy to
the gas \citep{whalen06}, so the photoionization heating rate is
\begin{equation}
\calg_{\rm ph} = \eGamma \sigma n_H \sum_n
\frac{s_n}{4\pi\left|\vecx-\vecx_n\right|^2} e^{-\tau(\vecx,\vecx_n)}.
\end{equation}
Each recombination of a hydrogen atom allows photons to escape,
leading to a loss of energy. A microphysical calculation of the
resulting cooling rate gives $\call_{\rm rec} \approx \Lambda_{\rm rec}
n_e \nHplus$, where \citep{osterbrock89}
\begin{equation}
\label{recombcool}
\Lambda_{\rm rec} \approx \left(6.1\times 10^{-10} \mbox{
cm}^{3}\mbox{ s}^{-1}\right) k_B T
\left(\frac{T}{\mbox{K}}\right)^{-0.89} 
\end{equation}
for temperatures $T \gtsim 100$ K, and recombination cooling falls to
negligible values at lower temperatures.

For heating and cooling that are not directly due to ionization of
hydrogen, we use simple optically-thin heating and cooling curves. In
molecular gas, we adopt the approximate cooling and heating functions
of \citet{koyama02}, and in partially ionized gas we compute the
cooling rate following \citet{osterbrock89}. Our calculation includes
the cooling by ion-electron collisions involving the first and second
ionized states of O, N, and Ne, which are the dominant coolants in HII
regions under solar metallicity conditions. We also include free-free
cooling. Thus, the total heating and cooling rates are
\begin{equation}
\label{otheating}
\calg = \eGamma \sigma n_H \sum_n
\frac{s_n}{4\pi\left|\vecx-\vecx_n\right|^2} e^{-\tau(\vecx,\vecx_n)}
+ n_H \Gamma_{\rm KI}
\end{equation}
and
\begin{equation}
\label{otcooling}
\call = \Lambda_{\rm
KI}(T) n_H^2 + \Lambda_{\rm rec}(T) n_e \nHplus +
\Lambda_{\rm ion-ff}(T) n_e \nHplus
\end{equation}
where the subscripts KI, rec, and ion-ff indicate heating and cooling
following the \citeauthor{koyama02} curves, from recombinations, and from
ion-neutral collisions and free-free emission in ionized gas.
With these cooling curves, the equilibrium temperature in neutral gas
at a density of $n_H=100$ cm$^{-3}$ is approximately $11$ K, and the
equlibrium temperature in fully ionized gas is approximately $6400$ K,
with some weak density dependence.

\subsection{Atomic Gas and PDRs}
\label{PDRs}

Here we justify our earlier statement that we can neglect PDRs in
studying the large-scale dynamics of most HII regions. We do this
using the analytic PDR models of \citet{bertoldi96}. Note that
\citeauthor{bertoldi96} consider the case of convex fronts, in which a
neutral cloud is bathed in ionizing radiation, whereas we are more
concerned with concave fronts, for which the source is embedded within
a molecular medium. To avoid complications arising from this, we will
only consider the case of plane-parallel ionization fronts. However,
the results should qualitatively generalize to our concave case.

First consider a system a short time after an ionizing source turns
on, before a shock front forms. In this case, there is only a
dissociation front and an ionization front. The propogation speed of
each front is then determined solely by the rate at which ionizing or
dissociating photons reach it. Let $s_{\rm Ly}$ and $s_{\rm
FUV}$ be the stellar luminosities in units of ionizing Lyman continuum
($\lambda < 912\,\AA$) and dissociating FUV
($912\,\AA<\lambda<1110\AA$) photons per second emitted by the
ionizing source, and $\tau_{\rm Ly}$ and $\tau_{\rm FUV}$ be the
optical depths from the source to the respective fronts. In this case
an ionization front at a distance $r_{i}$ from the source
propogates at a velocity
\begin{equation}
v_{i} = \frac{s_{\rm Ly} e^{-\tau_{\rm Ly}}}{4\pi r_i^2
n_H},
\end{equation}
where $n_H$ is the number density of hydrogen nuclei outside the
front. The dissociation front propogates at a rate
\begin{equation}
v_{d} \approx 0.15 \frac{s_{\rm FUV} e^{-\tau_{\rm FUV}}}{4\pi
r_{d}^2 n_H},
\end{equation}
where $r_d$ is the front radius, and the factor of 0.15 arises because
only a minority of FUV photons actually produce dissociations. For
stars of type O3 or earlier, i.e. those that drive significant HII
regions, $s_{\rm Ly}/s_{\rm FUV} \approx 1-2$, so $v_i > v_d$ unless
the Lyman continuum optical depth is large enough compared to the FUV
optical depth so that the fraction of Lyman continuum photons reaching
the front is $\approx 8-15\%$ of the number of FUV photons reaching the
front. This requires $\tau_{\rm Ly} - \tau_{\rm FUV} \gtsim 1.9-2.6$.

Since the dust extinction cross section for FUV photons is larger than
for Lyman continuum photons by a factor of $\approx 1.25$
\citep{bertoldi96}, this can only happen if there is significant
attenuation of Lyman continuum photons due to recombining gas within
the HII region. At early times before recombinations are significant
compared to photoionizations, this means $v_i > v_d$ and
the ionization and dissocation fronts will be coincident. In this
case, our neglect of the dissocation front is obviously not a
problem.

As the ionization front radius approaches the Str\"omgren radius,
\begin{equation}
r_s = \left(\frac{3 s_{\rm Ly}}{4\pi \alphaB n_H^2}\right)^{1/3},
\end{equation}
recombinations within the ionized region produce a significant neutral
population that raises $\tau_{\rm Ly}$ but not $\tau_{\rm FUV}$ and slows
expansion of the ionization front relative to the dissociation
front. If this process continued indefinitely then the dissocation
front would eventually race ahead of the ionization front. However,
slowing of the ionization front will also
eventually allow a shock to form, converting it from R type to D type,
and this will sweep gas up into a
dense shell and modify the propagation speed of both the ionization
and dissociation fronts. How much can the ionization front slow down
before a shock forms? The condition for a shock to form is that
the ionization front speed drops to of order the speed of a strong
shock driven by the ionized gas, which is twice the ionized gas sound
speed $c_i$. Thus, a shock forms when
\begin{equation}
\label{shockcondition}
\frac{s_{\rm Ly} e^{-\tau_{\rm Ly}}}{4\pi r_i^2 n_H} \approx 2 c_i.
\end{equation}
If we now approximate that $r_i \approx r_s$ once $\tau_{\rm Ly}
\gtsim 1$, it follows that
\begin{eqnarray}
\tau_{\rm Ly} & \approx & \frac{1}{3} \ln \left(
\frac{s_{\rm Ly} \alpha^{(B)^2} n_H}{96 \pi c_i^3}\right) \\
& = & 4.1 \ln \left(s_{49} n_2^2 \alphaB_4 c_6^{-3}\right),
\end{eqnarray}
where in the second step we have scaled to typical values for Galactic
molecular clouds and HII regions, $s_{49} = s_{\rm Ly}/(10^{49}\mbox{
s}^{-1})$, $n_2 = n_H/(100\mbox{ cm}^{-3})$, $\alphaB_4 = \alphaB /
\alphaB_{T=10^4\mbox{ K}} = \alphaB / (2.59\times 10^{-13}\mbox{
cm}^{-3}\mbox{ s}^{-1})$, and $c_6 = c_i/(10^6\mbox{ cm
s}^{-1})$. Thus, the dissociation front will not begin moving faster
than the ionization front until $\tau_{\rm Ly} \gtsim 1.9-2.6$, but a
shock will form once $\tau_{\rm Ly} \approx 4$.

Once a shock forms, we must consider how the width of the PDR compares to
the width of the shocked layer to determine if the dissociation front
can ever propogate ahead of the shock. In equilibrium, the column
density of hydrogen atoms through a PDR is roughly \citep{bertoldi96}
\begin{equation}
N_{H} \approx \sigma_{\rm FUV}^{-1} [1 + (2.7+\phi_0^{-1})^{4/3}]^{-1},
\end{equation}
where $\sigma_{\rm FUV}\approx 7.6\times 10^{-22}$ cm$^{-2}$ is the
attenuation cross section of dusty gas to FUV photons and the factor
$\phi_0$ is a dimensionless number describing the balance between
H$_2$ formation and dissociation. Since the numerical factor involving
$\phi_0$ is at most 0.21, this implies that there is a maximum
hydrogen column $N_{\rm PDR} \approx 2.8\times 10^{20}$ through a
PDR. Non-equilibrium PDRs are generally narrower than this. The column
of hydrogen swept up in the dense shell around an ionized region
shortly after it forms is roughly $n_H r_s$.
If we compare this to the maximum column density of a PDR, we find
that the column of hydrogen $n_H r_s$ present when the shock
first forms exceeds the maximum column of a PDR whenever
\begin{eqnarray}
s_{\rm Ly} & \gtsim & \frac{4\pi}{3} \alphaB \frac{N_{\rm PDR}^3}{n_H}
\\
& = & 2.3\times 10^{47} \alphaB_4 n_2^{-1}
\mbox{ s}^{-1}.
\end{eqnarray}
Thus, for $s_{\rm Ly} \gtsim 10^{47}$ s$^{-1}$, which is the case for
any strong ionizing source, the dissociation front will essentially
always remain trapped between the ionization front and the shock
front, except perhaps for a very brief interval when the front is changing
from R type to D type and the Lyman continuum optical depth through
the ionized region is about $2-4$. As a result, the dissociation front
will not have a significant effect on the dynamics of the ionized
region, since the only gas it can affect will be gas that has already
been shocked and is about to be ionized. The dissociation front will
simply be passively carried along. 

This can break down if we consider weak ionizing
sources, or if we have a medium substantially less dense than $n_H =
100$ cm$^{-3}$, the typical mean density in a GMC. If either of these is
the case, then a dissocation front may run ahead of the shock front
and pre-heat the gas, thereby affecting the dynamics. In this case our
neglect of the atomic gas becomes more problematic. However, even when
pre-heating occurs, the dissociation front will only precede the shock
front for a limited time, until the ionized region has swept up the
critical column density $N_{\rm PDR}$. Thus, even for weak ionizing
sources or low density regions, our approximation will only be invalid
around the time when the front transitions from R type to D
type.

Our simple analytic calculation is in good agreement with the detailed one-dimensional simulations of \citet{hosokawa05}, who also find that the molecular hydrogen dissociation front around an expanding HII region becomes trapped between the shock and ionization fronts shortly after the front transitions from R type to D type.

\section{Simulation Algorithm}
\label{algorithm}

\subsection{Program Flow}

We solve the evolution equations on a Cartesian grid with uniform cell
spacing $\Delta x$ using an operator split approach, in
which we alternate conservative MHD updates with the source terms on
the right hand sides set to zero with radiation updates in which we
add the source terms. In designing this update algorithm, we bear in
mind two factors. First, due to the comparitively simple nature of
the radiation update, computing a single explicit update using the
radiation source terms is computationally much cheaper than computing
a single magnetohyrodynamic update. For this reason, we choose to
sub-cycle the radiation update relative to the MHD update. This
effectively makes our radiation update step implicit, since during any
given simulation time step we iterate the temperature and chemical
state of each cell over many explicit update cycles.

Second, the overall
update time step is restricted by both radiation and MHD. The time
step must obey the Courant condition, but we also wish to impose a
heating/cooling time step condition. If the energy in a cell changes
too much between MHD updates, the corresponding large changes in
pressure from one time step to the next may lead to inaccurate
solutions. We therefore wish to limit the amount by which the energy
in a cell can change between MHD updates, which in turn imposes a time
step constraint associated with the rate of heating or cooling. Note
that, in contrast, the MHD update is not directly affected by the
neutral density $\rhon$, so we need not impose a corresponding
constraint on the overall simulation time step arising from changes in
the chemical state of the gas.

\begin{figure}
\plotone{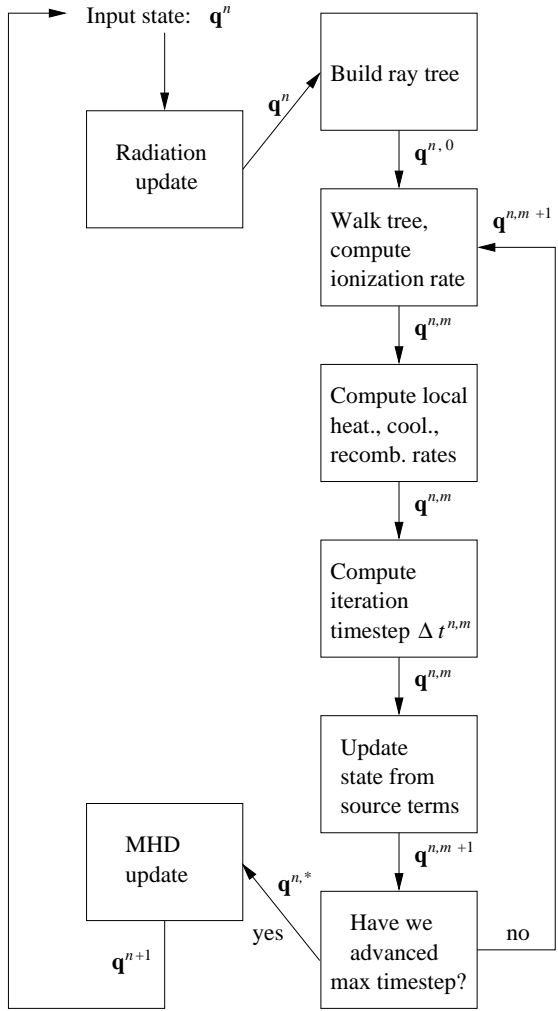}
\caption{
\label{progflow}
Schematic diagram of the main update loop of our method.
}
\end{figure}

In order to satisfy these constraints, we perform a two-step update
procedure, illustrated schematically in Figure \ref{progflow}. At the
end of time step $n$, in each cell we have a vector of quantities
\begin{equation}
\vecq^{n} = \left(
\begin{array}{c}
\rho^{n} \\
\vecp^{n} \\
\vecB^{n} \\
E^{n} \\
\rhon^{n} \\
\end{array}
\right)
\end{equation}
at time $t^{n}$. (Here $\vecp$ is the gas momentum, and we are
glossing over the fact that in the Athena algorithm the magnetic field
is actually stored at cell faces rather than cell centers.) We wish to
determine the properties $\vecq^{n+1}$ at time $t^{n+1}$.
We first update the gas energy $E$ and neutral
density $\rhon$ using only the terms on the right hand sides of
equations (\ref{massconservation}) - (\ref{neutralconservation}). This
update covers a time $\Delta t^n = \min(\Delta t_C, \Delta
t_{\epsilon})$, where $\Delta t_C$ is the Courant time step for the
current configuration, and $\Delta t_{\epsilon}$ is a time step
imposed by a requirement that the internal energy of a cell not change
by too much per time step. This gives us an intermediate
vector $\vecq^{n,*}$. Since we do not want to have to perform an MHD
update every time we do a radiation update, the radiation update
procedure is iterative. Therefore $E^{n,*}$ is not simply $E^{n} +
\Delta t^n (\calg - \call)^{n}$, and similarly for $\rhon^{n,*}$.  We
give a step-by-step description of the radiation update procedure in
\S~\ref{radiationalgorithm}, and defer formal definition of $\Delta
t_{\epsilon}$, $E^{n,*}$, and $\rhon^{n,*}$ until then.

Second, we perform an MHD update to gas quantities starting from state
$\vecq^{n,*}$, advancing the conservation equations
(\ref{massconservation}) - (\ref{neutralconservation}) with the source
terms on the right hand sides set to zero. This advance covers the
same time $\Delta t^n$ as the radiation update. We perform this update
using the standard Athena explicit MHD update procedure, which is
described in \citet{gardiner05, gardiner06}, with the trivial addition
of the passive scalar advection equation (\ref{neutralconservation})
for the neutral density. The overall program flow is independent of
the details of the MHD update algorithm. For the runs we report in
this paper, we configure Athena to use its directionally-unsplit
6-solve upwind constrained-transport method, the Roe Riemann solver with H correction,
and second-order spatial accuracy. We do not describe these methods in
greater detail
here, and instead refer readers to the Gardiner \& Stone papers.
At the end of this step, we have updated $\vecq^{n}$ for all the terms in
the evolution equations (\ref{massconservation}) -
(\ref{neutralconservation}), so the final state is $\vecq^{n+1}$, and we
may begin the next time step.

It should be noted that our update cycle is quite similar to that of
\citet{whalen06}.

\subsection{Radiation Update Method}
\label{radiationalgorithm}

Our radiation update algorithm combines elements of the approaches
\citet{abel02} and \citet{whalen06}, together with some novel
techniques that through the tests we describe in \S~\ref{tests} we
found necessary to achieve high accuracy. Our procedure to advance the
input state $\vecq^{n}$ to the intermediate state $\vecq^{n,*}$is as
follows. 

We do two prepatory steps before beginning a radiation
update. First, we compute the Courant time step $\Delta
t_C^n$ for the configuration $\vecq^n$. Second,
we construct a tree of rays following the prescription of
\citet{abel02} for each radiation source. We will not review the full
formalism of this method here, but a summary of its relevant
properties is that from each source one sends out a tree of rays in
directions that correspond to the centers of pixels in the HealPix
pixelization scheme \citep{gorski05}. At its coarsest level the tree
contains 12 rays uniformly spaced over the unit sphere, but the scheme
is hierarchical, such that any ray may be subdivided into four child
rays an arbitrary number of times. After the $l$th such division, the
unit sphere is discretized into $12(4^l)$ pixels, so that each pixel
covers a solid angle $\Omega = (\pi/3) 4^{-l}$ sr. As we follow rays
outward, we subdivide them to guarantee that at a distance $r$ from
the source $\Omega \le (1/2) \Delta x^2 / (4\pi r^2)$, i.e. that the
solid angle that a given ray represents is always smaller than the
solid angle subtended by a cell of the Cartesian grid by at least a
factor of 2. We always divide each ray at least twice, so we never
have fewer than $192$ rays.

Along each ray we construct a list of the cells through
which that ray passes, and the length of the ray segment that lies
within each cell. If we are running a calculation in parallel, so that
the computational domain is subdivided into a series of processor
domains that are contiguous rectangular blocks, we first construct the
list of processor domains through which a ray passes. On each
processor we only store information about rays for which either the
ray, its parent, or one of its children intersects that processor's
domain. This enables us to pass information from one processor to the
next as we walk outward along the ray tree without storing the full
tree on each processor.

If the source is moving, we repeat this procedure
every time the source moves, generally every time we perform an MHD
update. For a source that is fixed in space, we
build a new tree every fifth MHD time step. Each time we build the
tree we rotate it into a random orientation to minimize errors due to
discretization in angle. We discuss this procedure, and why it is
necessary, in more detail in \S~\ref{treerebuild}.

With these preparations done, we begin the iterative update
procedure. We compute the ionizing flux passing through
every cell by walking outward along the rays, following the procedure
outlined in \citet{abel02}. Suppose a particular ray representing a
solid angle $\Omega$ starts at a distance $r_0$ from the source, and
that the flux of ionizing photons at this point is $F_0$. The ray
passes through a sequence of $I$ cells, and we define $\ell_i$ as the
length of the ray segment that intersects the $i$th cell, $r_i = r_0 +
\sum_{j=0}^{i-1} \ell_i$ as the distance from the source to cell $i$,
and $F_i$ as the radiation flux reaching cell $i$. The optical depth
through cell $i$ is $\tau_i = (\sigma n_H + \sigma_d n) \ell_i$, where
$n_H$ and $n$ are the number densities of neutral hydrogen atoms and
hydrogen nuclei in the cell. The flux
reaching the next cell is just
\begin{equation}
F_{i+1} = F_i e^{-\tau_i} \left(\frac{r_{i+1}}{r_i}\right)^2,
\end{equation}
and conservation of ionizing photons therefore implies that the number
of ionizations per unit time in cell $i$ due to photons moving along
this ray is $F_i (1 - e^{-\tau_i}) \Omega r_i^2$. The corresponding
rates of mass photoionization per unit volume and photoionization
heating per unit volume are
\begin{eqnarray}
\label{iray}
\cali_{\rm ph} & = & F_i (1 - e^{-\tau_i}) \Omega r_i^2
\frac{\muH}{\Delta x^3} \\
\label{gray}
\calg_{\rm ph} & = & F_i (1 - e^{-\tau_i}) \Omega r_i^2
\frac{\eGamma}{\Delta x^3}.
\end{eqnarray}

We use this procedure to trace the photons along each ray, starting
from the 12 coarsest ones. Once we have reached the end of a ray, we
pass the flux $F_I$ that escapes the final cell on to its child
rays and repeat this procedure for them. We continue until we either
reach the edge of the computational domain or until such a small
fraction of the photons remain (we use $10^{-3}$) that it is no
longer worthwhile to continue following them. The total mass
photoionization rate $\cali_{\rm ph}^{n,m}$ and photoionization
heating rate $\calg_{\rm ph}^{n,m}$ in a cell at time $t^{n,m}$ is
simply the sum of the rates contributed by all the rays that intersect
it, as determined by equations (\ref{iray}) and (\ref{gray}), for the
gas properties at that time $\vecq^{n,m}$.

Once we have determined the rates of change of the neutral mass and
energy due to photoionization, we determine the rates due to
recombination, collisional ionization, and optically thin heating and
cooling using equations (\ref{recomb}), (\ref{collion}),
(\ref{recombcool}), (\ref{otheating}), and (\ref{otcooling}). These
calculations are purely local, and we simply perform them using the
current state of the gas $\vecq^{n,m}$. For reasons we discuss in
\S~\ref{mixedcells}, we set the heating and cooling rate due to molecular
transitions to zero in cells with ionization fractions between $1\%$ and
$99\%$. Adding the rates from local processes together with the
photoionization rates gives the total rate of change of the gas energy
$(\calg - \call)^{n,m}$ and the neutral density $(\calr-\cali)^{n,m}$
at time $t^{n,m}$.

Once we know the rates of change of the energy and neutral density, we
compute the time step for this iteration,
\begin{eqnarray}
\lefteqn{\Delta t^{n,m} = \frac{1}{10} \cdot}
\nonumber \\
& &
\max \left[
\frac{\epsilon^{n,m}}{(\calg - \call)^{n,m}},
\frac{\rhon^{n,m}}{(\calr - \cali)^{n,m}},
\frac{(\rho-\rhon)^{n,m}}{(\calr - \cali)^{n,m}}
\right],
\end{eqnarray}
where the maximum is over the three quantities in the parentheses
computed over all cells. Thus, the iteration time step is
chosen so that the maxmium fractional change in the internal energy
$\epsilon$, neutral density $\rhon$, or ion density $\rho-\rhon$ in
any cell is 10\%. We also want to ensure that the total radiation
time we advance is no larger than the Courant time step in the initial
configuration, so if $t^{n,m} + \Delta t^{n,m} > t_n + \Delta t_C^n$,
we set $\Delta t^{n,m} = t_n + \Delta t_C^n - t^{n,m}$.

Once we have fixed the time step $\Delta t^{n,m}$ for this iteration
of the radiation update, we actually perform the update to find the
new state
\begin{equation}
\vecq^{n,m+1} = \vecq^{n,m} + \Delta t^{n,m} \left(
\begin{array}{c}
0 \\
\mathbf{0} \\
\mathbf{0} \\
\calg - \call \\
\calr - \cali
\end{array}
\right)^{n,m}.
\end{equation}
We terminate iteration if one of the following conditions is
satisfied: (1) we have advanced over the total time determined by the
Courant time step in the initial configuration, i.e. $t^{n,m+1} = t^n
+ \Delta t_C^n$; (2) we have advanced over a total time that is equal
to or larger than the Courant time step in the \textit{current}
configuration, i.e. $t^{n,m+1} \ge t^n + \Delta t_C^{n,m}$; (3) the
internal energy of a cell has changed by more than a factor of $f$
from the its initial value, i.e. $\max(\epsilon^{n,m}/\epsilon^n,
\epsilon^{n}/\epsilon^{n,m}) > f$, where again the maximum is over
both quantities and over all cells. If none of these conditions are
met, we do another iteration, starting with a calculation of the
photoionization rate by traversing the tree.

The state when iteration terminates is the intermediate state we use
as an input to the conservative MHD update. Thus, if iteration
terminates after $M$ cycles, then
\begin{equation}
\vecq^{n,*} = \vecq^{n} + \sum_{m=0}^{M-1} \Delta t^{n,m}
\left(
\begin{array}{c}
0 \\
\mathbf{0} \\
\mathbf{0} \\
\calg - \call \\
\calr - \cali
\end{array}
\right)^{n,m}.
\end{equation}
The time step for the MHD update is then set to $\Delta t = t^{n,M} -
t^{n}$, the time for which we have computed the radiation update. If
the iteration terminates due to condition (1), this will be the
Courant time step of the initial state, $\Delta t = \Delta t_C^n$. If
it terminates due to condition (2), this will be $\Delta t \approx
\Delta t_C^{n,M}$, the Courant time step of the intermediate,
post-radiation state. If it terminates due to condition (3), the time
step will be $\Delta t = \Delta t_{\epsilon}$, the time it takes for
the internal energy to change by a factor $f$ in the cell that changes
by the largest factor. In practice, the third condition is usually the most
restrictive, since as the ionization front advances over a single
cell the internal energy of that cell changes by a factor of $\sim
10^3$, and we find that $f \ll 10^3$ is required to obtain good
solutions. We present tests to determine what value of $f$ produces
acceptable results for simple radiation hydrodynamics problems in
\S~\ref{timestep}.

\subsection{Cooling in Mixed Cells}
\label{mixedcells}

As part of our algorithm, we set $\Lambda_{\rm KI}$ and $\Gamma_{\rm
KI}$, the cooling and heating rates due to molecular processes, to
zero in cells with ionization fractions
between $1\%$ and $99\%$. We do this in order to prevent a problem with
overcooling in mixed ionization cells. To understand why this is
necessary, let us first consider what happens at an ionization front
in reality. In the fully molecular gas outside the front, heating due
to cosmic rays and interstellar UV and cooling due to molecular
emission balance and the gas is effectively isothermal; in the fully
ionized gas on the other side, heating by ionizing photons and cooling
by recombination and forbidden line emission also balance to keep the
gas isothermal. The ionized and molecular regions are separated by a
PDR and an ionization front. The true thickness of the ionization
front is of order the mean-free path of an ionizing photon, which in a
gas at a typical molecular cloud density $n_H = 100$ cm$^{-3}$ is
$(\sigma n_H)^{-1} \approx 100$ AU. Once an expanding ionization front
becomes D type and sweeps up a dense shell of material, the thickness
will be even smaller. The ionization fraction will be between $1\%$ and
$99\%$ only within this thin layer.  Because the hot ionized gas on one
side of the front and the cool molecular gas on the other side are
physically separated, and there is no mechanism to efficiently
transport heat across the front, gas that gains energy from ionizing
photons cannot lose it via molecular cooling.

Now, however, consider what happens in a simulation with finite
resolution. If one simulates a region that is parsecs or more in size,
the $\ltsim 100$ AU width of an ionization front will generally be
much smaller than the size of a computational cell. As a result, the
ionization front will be smeared out into a width comparable to cell
size, and this numerical mixing will supply a mechanism to transport
heat from ionized gas into molecular gas. This can lead to a dramatic
overestimate of the energy loss rate, because molecular emission is so
efficient. In effect, the rate limiting step in cooling will become
how rapidly energy can be transported into neutral gas by numerical
mixing. We demonstrate this effect in a real simulation in
\S~\ref{Dfrontcooling}.

Ideally one would solve this problem by either resolving the
width of the ionization front (or at least coming close enough to
render the overcooling region too small to do much damage), or using
an interface-tracking technique to follow the position of the front on
a subgrid scale. However, the former option is prohibitively expensive
in three dimensions (though it is feasible in two dimensions,
e.g. \citealt{arthur06}), and the latter is extremely cumbersome in a
three-dimensional simulation. Therefore we adopt a much simpler
solution: since we know physically that cells at an ionization front
should contribute negligibly to the overall cooling rate, we simply
set the molecular cooling rate in these cells to zero. This leads to a very
minor underestimate of the cooling rate, but the error is small
because in reality the front has a tiny volume and therefore should
contribute negligibly to the overall cooling rate. As we show in
\S~\ref{Dfrontcooling}, this procedure enables us to match analytic
solutions to very high precision, whereas if we leave the cooling on
in mixed cells we find large errors relative to analytic
solutions.

Our method does assume that the boundary between the
ionized and molecular gas can be described well as a single, simple
front. If  instead the molecular and ionized gas are mixed into a
two-phase medium with a complex interface that has structures small
compared to a computational cell (for example dense clumps of cold
molecular gas surrounded by a hot, ionized interclump medium), then our
approach of simply turning off molecular cooling will fail. In this
case, one would either need to resolve the complex interface, or use a
detailed subgrid model of heating and cooling to account for the
two-phase nature of the medium.

\section{Accuracy and Convergence Tests}
\label{tests}

Here we present tests of our code against analytic solutions. Since we
are not aware of any analytic solutions for ionizing radiative
transfer with magnetohydrodynamics in more than one dimension, all of
the tests we present in this section are hydrodynamic rather than
magnetohydrodynamic, i.e. $\vecB = 0$. We present magnetohydrodynamic
results in \S~\ref{MHDresults}.

\subsection{R Type Ionization Fronts without Recombination}
\label{norecomb}

A first, basic test of our implementation is the propogation of a
spherical ionization front in a medium in which there is no
hydrodynamic evolution and no recombinations. Physically, this is the
type of expansion that should occur at the very earliest stages of an
HII region's life, before there has been time for either hydrodynamic
motions or for a significant number of recombinations to occur. In
this case, a source with ionizing luminosity $s$ placed in a uniform
medium with density $\rho$ produces an ionized region whose
radius is simply set by equating the number of ionizing photons
produced up to a time $t$ with the number of hydrogen atoms within the
radius $r_i$. Thus, we should find
\begin{equation}
r_i = \left(\frac{3 \muH s}{4\pi\rho}\right)^{1/3} t^{1/3}.
\end{equation}

This problem is effectively a test of photon conservation in our
code. Since our radiation method is in principle exactly
photon-conserving, we should be able to compute the position of the
ionization front accurately to the resolution of our computational
grid. We preform a test with a source of ionizing luminosity
$s=4.0\times 10^{49}$ ionizing photons s$^{-1}$ placed in a uniform
medium of density $\rho=2.34 \times 10^{-22}$ g cm$^{-3}$ ($n_H = 100$
cm$^{-3}$) and an initial temperature of 20 K. The source is at the
center of a computational domain that runs from $-10$ pc to $10$ pc in
each direction. For this test we disable the MHD update in our code,
and set the rates for all radiative processes except photoionization
to zero. We use a resolution of $64^3$ cells.

\begin{figure}
\plotone{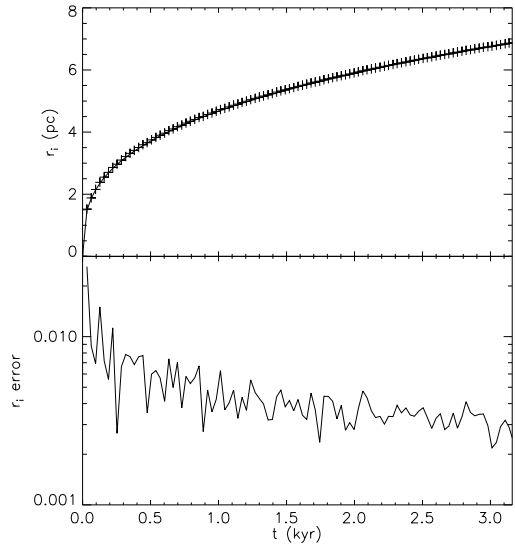}
\caption{
\label{rifnr}
Ionization front radius $r_i$ versus time $t$ (\textit{upper
panel}), and error in simulated $r_i$ relative to the analytic value
versus time (\textit{lower panel}) in a simulation of an R type
ionization front with no recombinations. The upper panel shows the radius
computed in the simulation (\textit{plus signs}) and the analytically
computed radius (\textit{solid line}). We define the ionization front
radius as the mean radius of the centers of all cells for which $0.01 <
\rhon/\rho < 0.99$. The error is defined as $|r_{\rm sim}-r_{\rm
analyt}|/r_{\rm analyt}$, where $r_{\rm sim}$ is the radius found in
the simulation and $r_{\rm analyt}$ is the analytically computed
value.
}
\end{figure}

\begin{figure}
\plotone{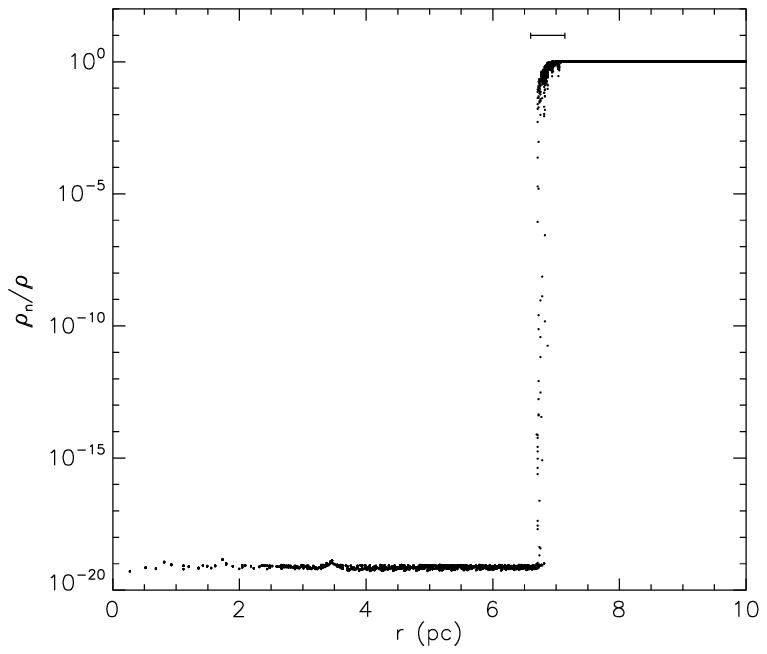}
\caption{
\label{rprofnr}
Ratio of neutral to total density $\rhon/\rho$ versus radius $r$ for
a subset of cells in our computational domain at a time of $10^{11}$ s
($3.17$ kyr) in a simulation of an R type ionization front without
recombinations. The scale bar above the location of the ionization front
shows a range of $\pm(\sqrt{3}/2)\Delta x$ about the analytically
computed radius of the ionization front at this time. The minimum
ratio of $10^{-20}$ in the ionized region is set by a
numerical floor in our code.\\
}
\end{figure}

\begin{figure}
\plotone{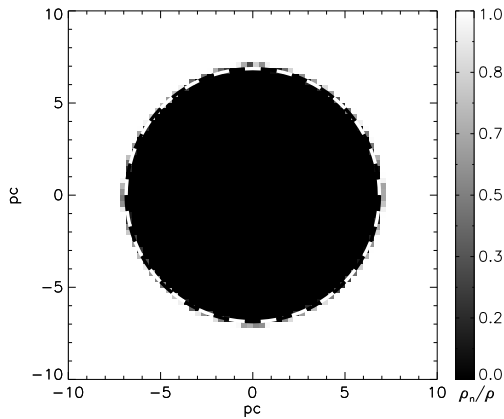}
\caption{
\label{rslicenr}
Ratio of neutral to total density $\rhon/\rho$ in an equatorial slice
through our 3D computational domain in a simulation of an $R$ type
ionization front without recombinations. The time shown is the same as
that in Figure \ref{rprofnr}. The dashed line shows the
analytically-computed front position.\\
}
\end{figure}

Figure \ref{rifnr} shows the computed radius of the ionization front
versus time in our test. As the plot shows, the computed solution
agrees with the analytic prediction to better than a percent, which is
an error much smaller than the size of a computational cell. Figure
\ref{rprofnr} shows the radial profile of the ionization fraction
versus radius for every cell in our computational domain at a time of
$10^{11}$ s ($3.17$ kyr), and Figure \ref{rslicenr} shows the neutral
fraction in an equatorial slice through the domain. As the plots show,
the ionization front is entirely confined to cells at the analytically
computed front radius at this time, $6.9$ pc. Thus, our code passes
this test quite well.

\subsection{R Type Ionization Fronts with Recombination}
\label{rtype}

A next test of our code is in the case where there are still no
motions, but recombinations do occur. Physically, this represents the
next phase in the expansion of an R type ionization front, when
the recombination rate becomes competitive with the photoionization
rate, but before the excess pressure in the ionized gas causes the HII
region to begin hydrodynamically expanding. If the recombination
coefficient $\alphaB$ is constant, this problem has an analytic
solution: at time $t$, the radius of the ionization front is
\begin{equation}
r_i = r_s \left(1-e^{-t/t_{\rm rec}}\right)^{1/3},
\end{equation}
where
\begin{equation}
r_s = \left(\frac{3s \muH^2}{4\pi \alphaB \rho^2}\right)^{1/3}
\end{equation}
is the Str\"omgren radius and
\begin{equation}
t_{\rm rec} = \frac{\muH}{\alphaB \rho}
\end{equation}
is the recombination time scale for the gas. At early
times this solution matches the case of powerlaw expansion discussed
in \S~\ref{norecomb}, but as times $t\gtsim t_{\rm rec}$ the
ionization front slows down due to recombinations. At times $t \gg
t_{\rm rec}$, the front radius approaches $r_s$ and the front velocity
drops to zero.

For any realistic cooling curve, for which the heaitng and cooling
rates depend on the density, radiation flux, and the ionization state,
there will be some small temperature variation in time and space
within the HII region, so $\alphaB$ will not be exactly constant and
the analytic solution will not apply exactly. To enable us to compare
with the analytic solution as well as possible, and therefore to make
the strongest possible evaluation of the code, for the purposes of
this test we set $\alphaB=2.59\times 10^{-13}$ cm$^{-3}$ s$^{-1}$
independent of temperature.

\begin{figure}
\plotone{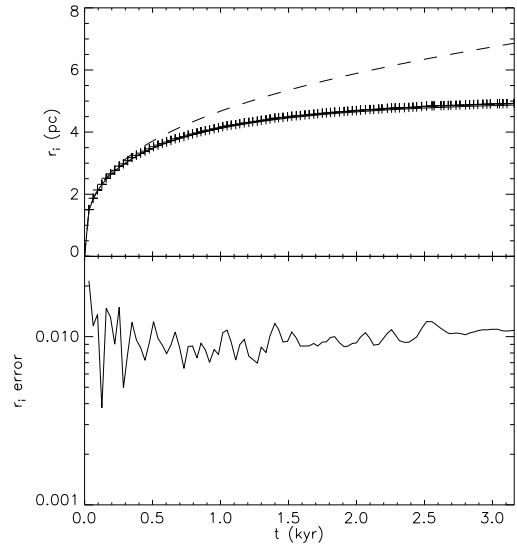}
\caption{
\label{rifnh}
Ionization front radius $r_i$ versus time $t$ (\textit{upper
panel}), and error in simulated $r_i$ relative to the analytic value
versus time (\textit{lower panel}) in a simulation of an R type
ionization front with recombinations. The upper panel shows the radius
computed in the simulation (\textit{plus signs}), the analytically
computed radius (\textit{solid line}), and the analytic prediction for
the radius if there are no recombinations (\textit{dashed line}). We
define the ionization front radius as the mean radius of the centers
of all cells for which $0.01 < \rhon/\rho < 0.99$. The error is defined
as $|r_{\rm sim}-r_{\rm analyt}|/r_{\rm analyt}$, where $r_{\rm sim}$
is the radius found in the simulation and $r_{\rm analyt}$ is the
analytically computed value.
}
\end{figure}

\begin{figure}
\plotone{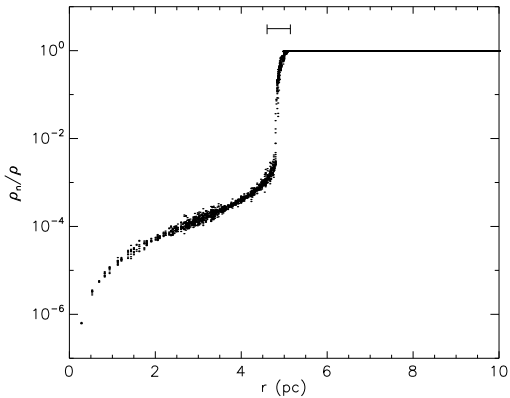}
\caption{
\label{rprofnh}
Ratio of neutral to total density $\rhon/\rho$ versus radius $r$ for
a subsample of cells in our computational domain at a time of $10^{11}$ s
($3.17$ kyr) in a simulation of an R type ionization front with
recombinations. The scale bar above the location of the ionization
front shows a range of $\pm(\sqrt{3}/2)\Delta x$ about the analytically
computed radius of the ionization front at this time.
}
\end{figure}

We run this test using the same initial conditions and resolution as
in \S~\ref{norecomb}, which give $r_s = 5.0$ pc and $t_{\rm rec} =
1.2$ kyr for our chosen value of $\alphaB$. For this test, we
again disable the MHD update step in our code, but allow the entire
radiation module to operate normally. We run for $10^{11}$ s
($3.17$ kyr), which is more than $2 t_{\rm rec}$. Figure \ref{rifnh}
shows the position of the ionization front versus time, computed as in
\S~\ref{norecomb}, and Figure \ref{rprofnh} shows the radial profile
of the neutral density at a time of $10^{11}$ s. As we found in
\S~\ref{norecomb}, the ionization front is entirely confined to a
radial extent of a single cell, so the front is as spherical as
possible given our Cartesian grid. The accuracy of the front position
is $\sim 1\%$, better than a single cell.

\subsection{D Type Ionization Fronts}
\label{Dfronttest}

The tests described in \S~\ref{norecomb} and \S~\ref{rtype}
demonstrate that our radiation module produces correct results on its
own, and the tests in \citet{gardiner05,gardiner06} show that the MHD
module works well. However, we must also test how the two modules work
together. We therefore consider the problem of the hydrodynamic
evolution of a uniform neutral medium with initial density $\rho_n$
and temperature $T_n$ within which there is an ionizing
source. This problem has a well-known solution. At early times there
is an R type ionization front: the source ionizes the gas out
to a radius $r_i\approx r_s$, and heats it to $T_i \approx 6400$ K,
but the gas does not move and the ionization front is not preceeded by
a shock. However, the ionized region is
overpressured relative to the surrounding gas, and it expands on a
sound crossing time scale, $t_s \equiv r_s/c_i$, where $c_i \approx
8\times 10^5$ cm s$^{-1}$ is the sound speed in the ionized gas. As
it expands, it sweeps up a dense shell of neutral material, forming a
D type ionization front. The expansion is subsonic with respect to the
ionized gas, so the ionized region has an almost uniform density and
temperature. Ionization balance requires that this density vary as 
$\rho_i\propto r_i^{-3/2}$, so
the shell is driven outward by a pressure $p=\rho_i
c_i^2\propto r_i^{-3/2}$. In the limit where the pressure in the
ionized region that is driving the shell greatly
exceeds the thermal pressure in the ambient medium, and the neutral
mass in the shell greatly exceeds the mass within the ionized region,
conservation of momentum for the shell implies that after a time $t$
it expands to a radius \citep{spitzer78}
\begin{equation}
r_{\rm sh} = r_s \left(1 + \frac{7 t}{4 t_s}\right)^{4/7}.
\end{equation}
The thickness of the shell relative to its radius is of order
$(c_n/v_{\rm sh})^2$, where $c_n$ is the isothermal sound speed in the
neutral gas and $v_{\rm sh}=c_i [1+(7/4)t/t_s]^{-3/7}$ is the velocity
of the shell. As with the analytic solution for the R type ionization
front, this strictly applies only if the ionized gas sound speed $c_i$
is constant. For a realistic cooling curve, we expect this to be
nearly but not perfectly true.

We run simulations with an initial density $\rho=2.34\times
10^{-22}$ g cm$^{-3}$, temperature $T=11$ K, and an ionizing source of
luminosity $s=4.0\times 10^{46}$ s$^{-1}$. The source is placed at the
center of a box running from $-10$ pc to $10$
pc in every direction. For these values of $\rho$ and $s$, and
assuming an ionized region temperature of roughly $8.0\times 10^3$ K,
we have $r_s = 0.5$ pc and $t_s=0.061$ Myr. We expect that the similarity
solution will apply for times $1 \ll t/t_s \ll 1000$. The lower limit on
the time is imposed by the requirement that the mass in the swept-up
shell be much larger than the mass in the interior of the HII region,
and the upper time limit comes from the requirement that the pressure
in the HII region greatly exceed the ambient pressure, or equivalently
that the thickness of the shell be much smaller than its radius. We
run all our tests for 1 Myr, or $16.3 t_s$.

\subsubsection{Comparison to the Analytic Solution}

\begin{figure*}
\plotone{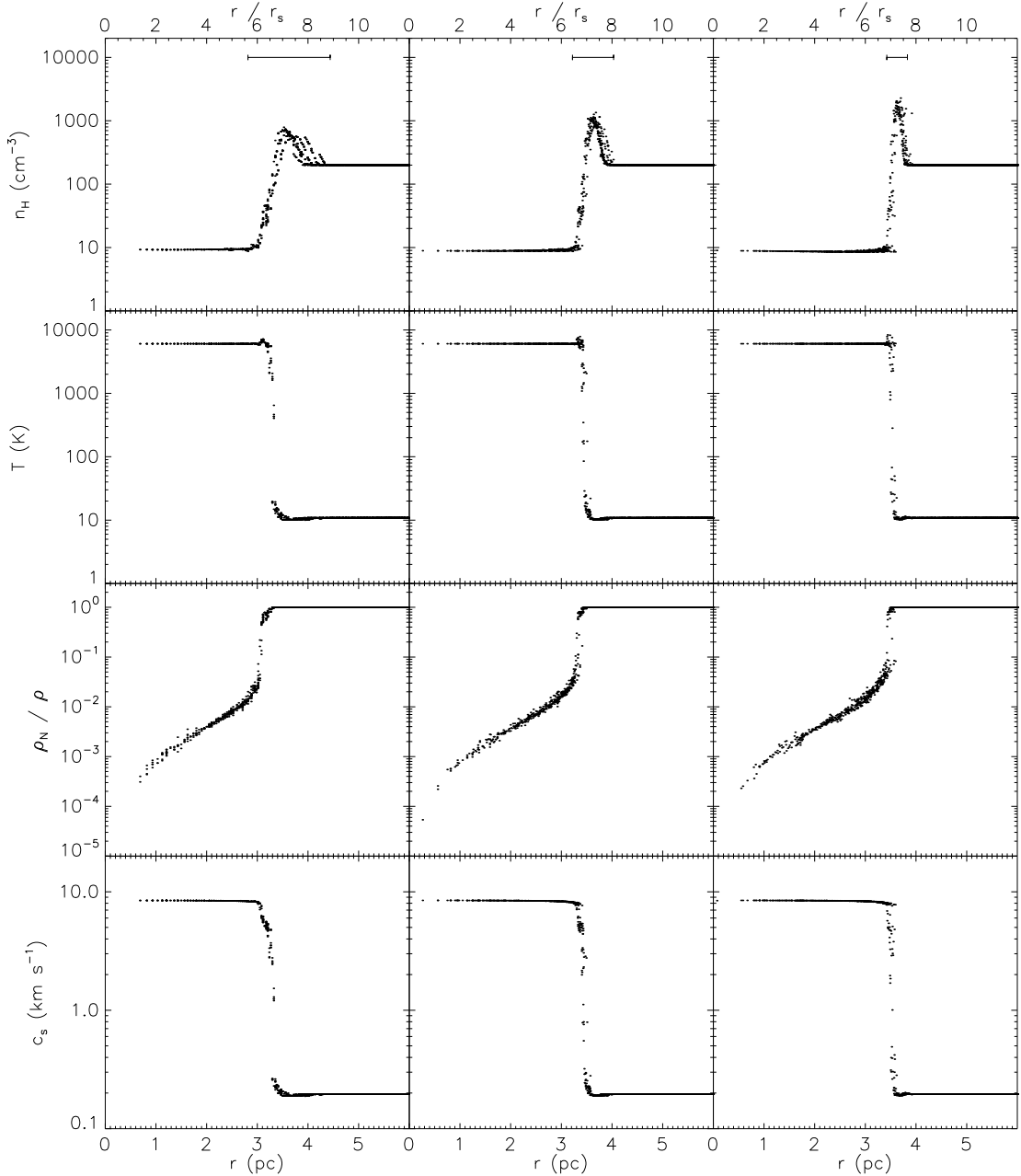}
\caption{
\label{rprofd}
Hydrogen number density, temperature, neutral fraction, and isothermal
sound speed versus radius for a selection of computational cells in
our simulation of a D type ionization front, at resolutions of $64^3$
(\textit{left column}), $128^3$ (\textit{middle column}), and $256^3$
(\textit{right column}). We show only a subsample of the gridpoints to
minimize clutter and file size. The points are plotted at a time of
$1$ Myr ($17.7 t_s$) after the start of the
simulation. The scale bars in the uppermost panels indicate a range of
$\pm 3 (\sqrt{3}/2)\Delta x$ about the analytically computed radius of
the dense shell at this time, to indicate a width of 3 cells, about
the expected width of a shock due to numerical viscosity.\\
}
\end{figure*}

\begin{figure}
\plotone{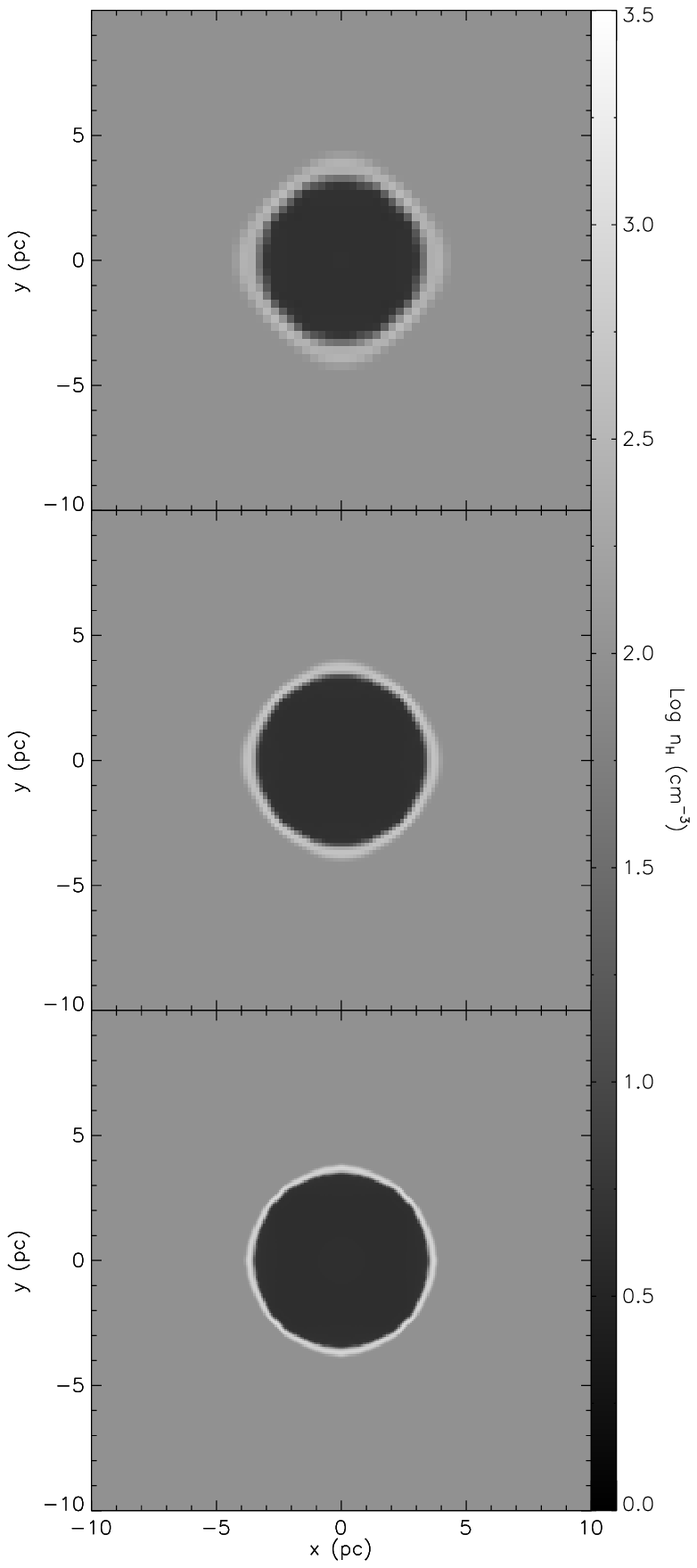}
\caption{
\label{dimg}
Slice through the $xy$ plane showing log of hydrogen number
density in a simulation of a D type ionization front 1 Myr ($17.7t_s$)
after the start of the simulation, at a resolution
of $64^3$ (\textit{top panel}), $128^3$ (\textit{middle
panel}), and $256^3$ (\textit{bottom panel}).
The slices in the $xz$ and $yz$ planes are nearly identical.\\
}
\end{figure}

We begin with simulations at resolutions of $64^3$, $128^3$, and
$256^3$, using a factor $f=4$ for our heating time step constraint
(i.e. we allow the
internal energy of a cell to change by up to a factor of 4 between
hydrodynamic updates.) 
Figure \ref{rprofd} shows the radial distribution of densities,
temperatures, neutral fractions, and isothermal sound speeds in the
gas in the simulations after 1 Myr. As we expect, the density and
temperature
in the ionized region are almost constant. The density peaks at the
position of the overdense shell, but the shell is spread over a few
cells, so its width decreases with resolution. This
spreading is to be expected. The true physical thickness of the shell
at this time should be roughly a percent of $r_{\rm sh}$,
smaller than the size of a single computational cell. However, our
code has some numerical viscosity, particularly for motions
that are not aligned with the computational grid, and
this spreads shocks over $\sim 3$ cells. As Figure \ref{dimg} shows,
despite this spreading the front remains round at each resolution.

\begin{figure}
\plotone{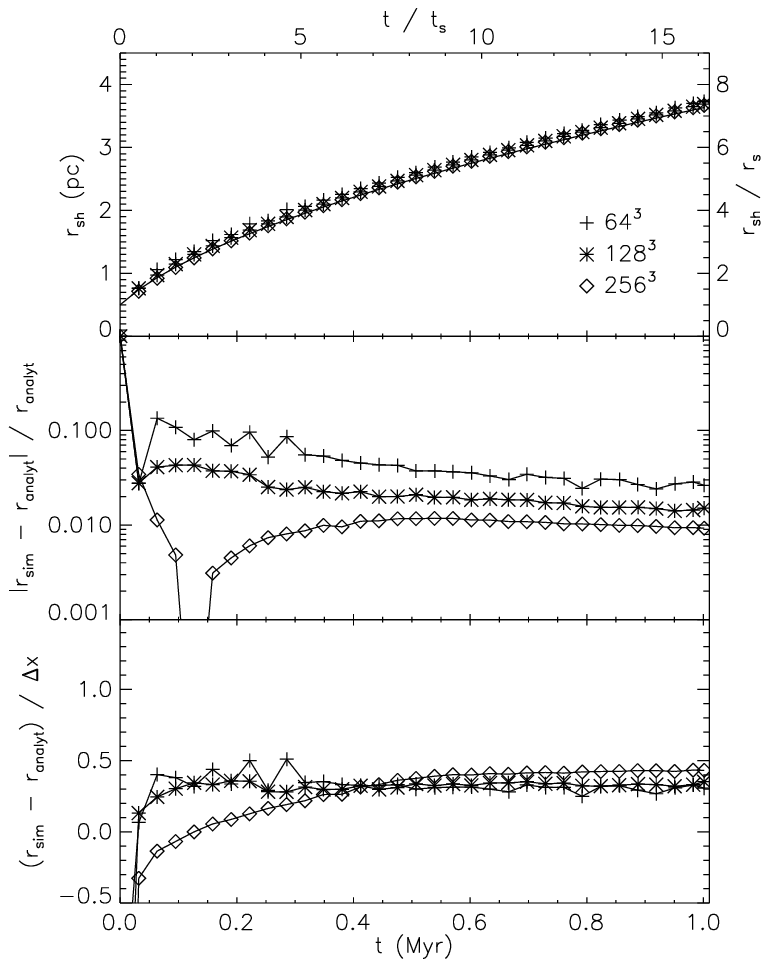}
\caption{
\label{rdfres}
Radius of the dense shell around an HII region verus time
(\textit{top panel}), computed analytically (\textit{solid line})
and by simulations at resolutions of $64^3$ (\textit{plus signs}),
$128^3$ (\textit{asterisks}), and $256^3$ (\textit{diamonds}).
We also show the difference between the simulated and
analytically-computed shell radii, $r_{\rm sim}$ and $r_{\rm
analyt}$, normalized to the analytic radius (\textit{middle panel})
and normalized to the size of a computational cell (\textit{bottom
panel}).
}
\end{figure}

We next compute the radius of the dense shell as a function of time in our
simulation. We do this by identifying all the cells whose densities are at
least 10\% larger than the initial density, and then taking the average of their radii. 
Using an overdensity cutoff different from $10\%$ modifies the
results in detail, but not qualitatively. We compare this measurement
to the analytic solution in Figure \ref{rdfres}. As the plot shows,
the radius of our simulated shell matches the analytically computed
radius at late times quite well in all the runs. The error is always
smaller than cell in size, which is the best that can be expected using a
Cartesian grid with finite numerical viscosity. This error is $\sim
3\%$ of the analytically computed radius at the lowest resolution and
less than $1\%$ at the highest resolution. The error generally
decreases with time, both because the ratio of the radius to a cell
size is increasing, and because the analytic solution is becoming a
better and better approximation as time increases.

\subsubsection{Heating Time Step Convergence Tests}
\label{timestep}

We have shown that our method can match the analytic solution for the
propogation of a D type ionization front with a heating time step
constraint of $f=4$, meaning that we allow the pressure in a cell
to change by at most a factor of 4 between hydrodynamic
updates. However, we would like
to explore a range of heating time step constraints to determine if
the accuracy of the solution is affected by the choice of $f$. We
therefore re-run our $128^3$ simulation with $f=2$, $f=10$, and $f=100$ to
see if the quality of the solution changes significantly. The choice
$f=100$ corresponds to essentially no constraint on the hydrodynamic
time step other than the ordinary Courant condition. We wish to
use as large a value of $f$ as possible, because that will minimize
the computational cost by enabling us to take fewer of the more
expensive MHD updates per radiation update. Of course if $f$ is too
large then the cost of the radiation update will dominate the total
computation cost, and there will be no additional advantage in increasing
$f$. Where exactly this crossover occurs depends on the number of
processors and the geometry of the ionization front.

\begin{figure}
\plotone{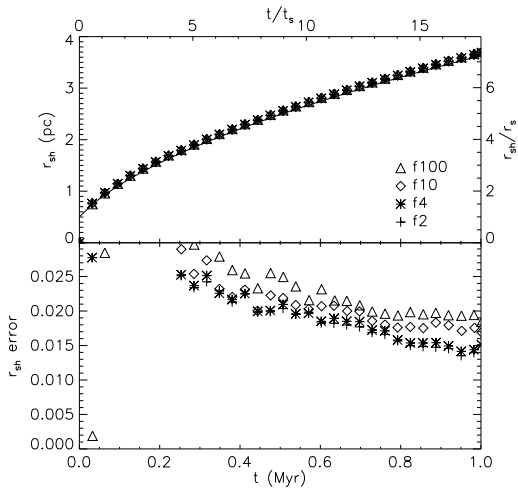}
\caption{
\label{rdff}
Radius of the dense shell around an HII region verus time
(\textit{upper panel}), computed analytically (\textit{solid line})
and by simulations at a resolution of $128^3$, with $f=2$
(\textit{plus signs}), $f=4$ (\textit{asterisks}), $f=10$
(\textit{diamonds}), and $f=100$ (\textit{triangles}).
We also show the fractional error in the
position of the shell relative to the similarity solution
(\textit{lower panel}), defined as $|r_{\rm sim} -
r_{\rm analyt}|/r_{\rm analyt}$, where $r_{\rm sim}$ is the simulated
radius and $r_{\rm analyt}$ is the analytically computed radius.
}
\end{figure}

Figure \ref{rdff} shows the radius of the shell versus time as
computed in our simulations with varying values of $f$, and as
computed analytically. The results show that the radius approaches the
analytic solution somewhat fall all $f$, but that the error is larger
and the convergence slower for larger values of $f$.
This suggests that numerical methods in which the
temperature change between hydrodynamic updates is unconstrained
\citep[e.g.][]{dale05, mellema05, mellema06} or is limited only to
values of $f\gg 10$ \citep[e.g.][]{maclow06} may produce
quantitatively incorrect results for the expansion rates of D type
ionization fronts, although the error is likely to be only a few
cells. This problem may not affect methods that implicitly update
the hydrodynamics and the radiation together
\citep[e.g.][]{miniati06}, rather than in an operator split fashion,
but we are unaware of any such methods for ionizing radiation
transport as opposed to local heating and cooling functions.

Note, however, that even with large $f$ the error does decrease in
time. This result is easy to understand intuitively. Once the front
is well into the D type phase, the Courant time step is set by the
sound speed in the ionized region, which is roughly constant. As the
front sweeps up more material and slows down, the rate at which gas is
heated from the inital temperature to the ionized temperature
decreases, so, on average, the fractional change in cell temperature
per fixed Courant time step is a decreasing function of time. Thus,
the error one makes by allowing a large change in gas temperature per
Courant time step decreases with time. This means that codes with
unconstrained temperature changes per hydrodynamic update are likely
to be most inaccurate at early times when fronts are moving at speeds
close to the ionized sound speed, and are most accurate at late times
when fronts move slowly.

\subsubsection{Ray Rotation}
\label{treerebuild}

\begin{figure}
\plotone{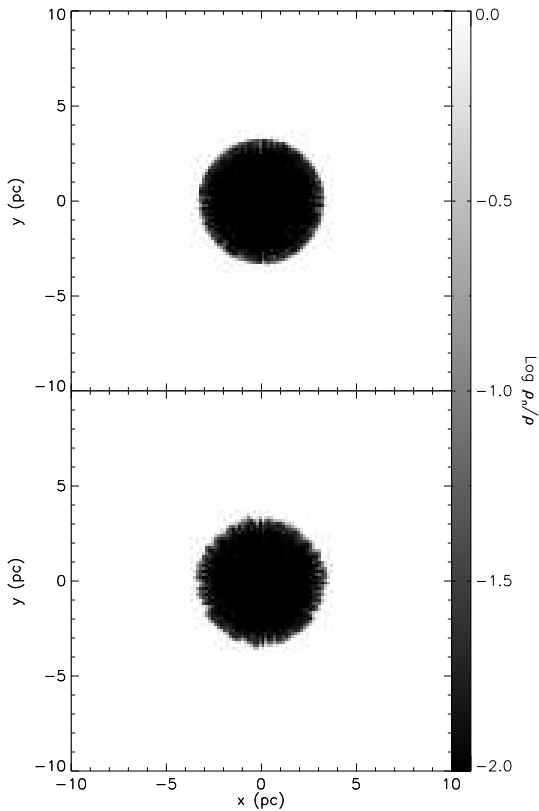}
\caption{
\label{rebimg}
Slice through the $xy$ plane showing log of the ratio of neutral
density to total density in a simulation of a D type ionization front
1 Myr ($17.7t_s$) after the start of the simulation, with
(\textit{upper panel}) and withpout (\textit{lower panel}) rotation of
the rays.
}
\end{figure}

To demonstrate why it is necessary to rotate the orientation of the
rays in our calculation regularly, we repeat our $128^3$, $f=4$ test
calculation without ever rotating the rays. The mean radial positions
of the dense shell and the ionization front are essentially the same
in the two calculations, but the shape of the ionization front
is significantly different. Figure \ref{rebimg} shows
the neutral fraction in slices through the simulations without and
with ray rotation after 1 Myr of evolution. With rotation of the rays,
as shown in the lower panel,
the ionization front boundary is smooth and round. Without rotation,
as shown in the upper panel,
there are projections of partially neutral gas several cells into the
ionized region, and the front boundary is much less smooth. If
one is interested in studying instabilities or similar phenomena at
the ionization front boundary, the presence of these fingers of
neutral gas is potentially problematic, since they may serve as seeds for
instability.

These projections are caused by differences in ionization rate for
cells at the same radius due to discretization in angle. When the
orientation of the rays is constant over long times, small differences
in ionization rate at different angles can build up to produce
the finger-like structures shown in the figure. One could reduce the
discretization error by using a more refined ray tree, for example by
requiring that the solid angle subtended by a cell always be at least
four times as large as the solid angle corresponding to a single ray,
rather than twice as large as we require. However, this is potentially
expensive, since every factor $f$ increase in the angular resolution
multiplies the time required to compute the photoionization rate by a
factor of $f^2$, and the photoionization rate is computed during every
radiation iteration. It is also overkill, since the angular
disretization errors only build up over many time steps. Rotating the rays
at intervals of 5 hydrodynamic time steps, as we have done for the
simulation shown in the lower panel of Figure \ref{rebimg}, eliminates
or vastly reduces the occurence of neutral gas fingers in the
ionized region. The computational cost of rebuilding the ray
tree every 5 hydrodynamic time steps is completely negligible and, for
parallel calculations, requires no additional inter-processor
communication.

\subsubsection{Ionization Front Cooling}
\label{Dfrontcooling}

We next explore the question of cooling in cells with mixed ionization
states. As discussed in \S~\ref{mixedcells}, we set the molecular
cooling rate $\Lambda_{\rm KI} = \Gamma_{\rm KI} = 0$ in
cells where the ionization fraction is between $1\%$ and $99\%$ in
order to prevent excessive cooling due to numerical mixing of ionized and
neutral gas. To demonstrate why this is necessary, we repeat our
simulation of a D type ionization front with $f=4$ with cooling
allowed in all cells regardless of ionization fraction.

\begin{figure}
\plotone{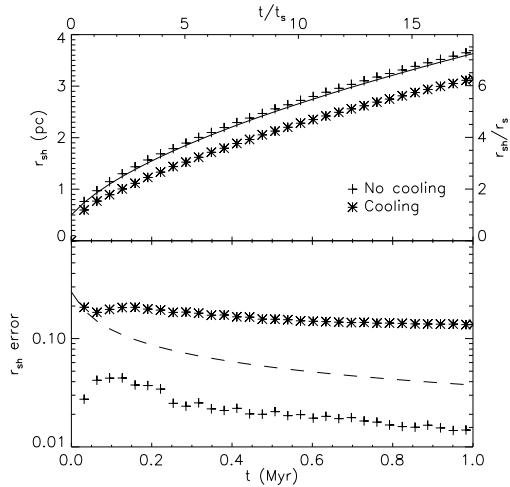}
\caption{
\label{rdfcool}
Radius of the dense shell around an HII region verus time
(\textit{upper panel}), computed analytically (\textit{solid line})
and by simulations at a resolution with cooling in mixed-ionization
cell disabled (\textit{plus signs}) and enabled
(\textit{asterisks}). We also show the fractional error in the
position of the shell relative to the similarity solution
(\textit{lower panel}), defined as $|r_{\rm sim} - r_{\rm
analyt}|/r_{\rm analyt}$, where $r_{\rm sim}$ is the simulated radius
and $r_{\rm analyt}$ is the analytically computed radius. For
comparison, we also show the fractional error corresponding to a
difference of $(\sqrt{3}/2) \Delta x$ (\textit{dashed line}), the
distance from the center of a computational cell to its corner.
}
\end{figure}

Figure \ref{rdfcool} shows the result. The run where all cells can
cool lags the analytic solution even at early times, and the error
does not decrease with time. The error at 1 Myr of evolution is
roughly 14\%, which is far larger than a single cell; the error
in the slope of the of the $r_{\rm sh}$ versus $t$ curve at 1 Myr is
10\%. In contrast, the error with mixed-ionization cooling disabled is
$1.5\%$, which corresponds roughly to the center-to-corner
distance of a single cell.  The slope of the curve differs from the
analytic value by less than 1\%.

\begin{figure}
\plotone{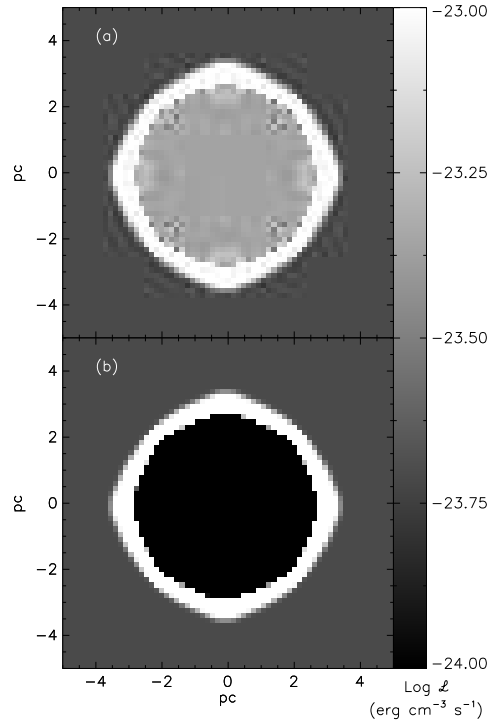}
\caption{
\label{coolrate}
Slice through the $xy$ plane showing the total cooling rate
(\textit{panel a}) and the cooling rate due to molecular processes
alone (\textit{panel b}). Note that, in order to make the cooling rate
in the fully ionized region visible, we have selected a color scale
that saturates in the rapidly cooling ionization front. The maximum
cooling rate reaches $>10^{-21}$ erg cm$^{-3}$ s$^{-1}$.
}
\end{figure}

The lag is a result of numerical mixing at the ionization front
coupled with strong molecular cooling. To demonstrate this, we plot in
Figure \ref{coolrate} both the total cooling rate $\call=\Lambda_{\rm
KI} n_H^2 + \Lambda_{\rm rec} n_e \nHplus + \Lambda_{\rm ion-ff} n_e
\nHplus$ (panel a) and the molecular cooling rate $\Lambda_{\rm
KI} n_H^2$ (panel b) as a function of position from our run with
cooling in mixed 
cells allowed, at a time of 1 Myr. As the plot shows, with cooling
allowed in mixed cells, the energy loss rate from cells at the
ionization front is much larger than that from cells either inside or
outside the front.

One might think
this is to be expected, since the cells at the ionization front are
the ones furthest out of equilibrium, and are the places where much of
the energy from ionizing photons is deposited. However, 91\% of the
cooling in the region with an ionization fraction larger than
$1\%$ comes from molecular cooling rather than from recombination
or forbidden line cooling, the processes that should dominate in
ionized regions. Molecular cooling dominates even if we consider only
much more strongly ionized regions. Only if we limit our attention to
gas where the ionization fraction is greater than $74\%$ do we find that
molecular processes provide less than 50\% of the total cooling
rate. That molecular cooling is the dominant cooling process even in
gas that is up to 74\% ionized is a clear sign that the cooling rate
is being artificially enhanced by numerical smearing of the ionization
front, since for a real ionization front there should be no molecules
to cool in regions that are significantly ionized. 

This suggests that
simulations can only
compute the correct expansion rate for HII regions if they suppress
the effects of numerical mixing at the ionization front. This may
explain why some numerical simulations
without this precaution produce results that lag the true solution
significantly \citep[e.g.][figure 1]{maclow06} unless they have enough
resolution to at least marginally resolve the ionization front
\citep[e.g.][]{arthur06}, something that is not generally feasible in
three dimensions. Note, however, that the strength of this effect
depends on the existence of extremely rapid cooling processes for the
non-ionized gas. If the chemistry is such that the neutral gas is not
able to cool rapidly, the error will be much smaller. For example,
with primordial cosmological chemistry the lag is much smaller because
the neutral gas cooling is much less efficient (D. Whalen, 2006,
private communication).

An important point is that the evolution in the case with overcooling
is \textit{qualitatively} correct. There is an expanding, spherical
HII region driving a dense shell of swept-up material, and following a
radius versus time curve that resembles the correct solution. It is
only when one \textit{quantitatively} checks the numerical method
against an analytic solution that the problem becomes clear. This
highlights the importance of performing quantitative comparisons of
numerical and analytic results.

\section{HII Region Evolution in Uniform Magnetized Media}
\label{MHDresults}

Now that we have determined the constraints our numerical method must
obey from tests against analytic solutions in the hydrodynamic case,
in this section we demonstrate our method in the simplest
magnetohydrodynamic case. Several authors have considred the structure
of magnetized ionization fronts in the past. However, these treatments
have been limited to mathematically rigorous but one-dimensional
analyses \citep[e.g.][]{redman98, williams00a,
williams01} or to simple, quasi-empirical models
\citep[e.g.][]{carlqvist03, ryutov05} in more
dimensions. \citet{bertoldi89} gives an analytic model for the
implosion of a spherical magnetized gas clump exposed to an external ionizing
source, but his solution is limited to that case. To date, there has
been no detailed simulation of the evolution of an HII region driven
by a point source expanding into a magnetized medium in three
dimensions.

We consider a neutral medium with initially
uniform density $\rho_n$ and isothermal sound speed $c_n$, threaded by a
magnetic field $\vecB_n$ oriented in the $x$ direction. We place a
source of constant ionizing luminosity $s$ at the origin that begins
radiating at time $t=0$. We restrict ourselves to the case most
relevant to ionizing sources in magnetized molecular clouds, in which
the thermal pressure in ionized gas at the initial density greatly
exceeds the magnetic pressure, which in turn greatly exceeds the
thermal pressure in neutral gas at the initial density. Thus, $c_i \gg
v_A \gg c_n$, where $v_A$ is the Alfven speed in the neutral gas and
$c_i$ is the ionized gas sound speed.

\subsection{Analytic Evaluation}

We can sketch out an approximate evolutionary scenario for this
configuration using simple analytic estimates. At times $t \ltsim
t_s$, the gas has not yet had time to move in response to
photoionization heating. Since photons do not feel the magnetic field,
the evolution should be identical to that of a non-magnetized R type
ionization front. At $t\sim t_s$, the ionized gas begins to expand due
to its high thermal pressure. Since by assumption $c_i \gg v_A$, the
magnetic field in the gas is initially unable to resist this
expansion. Thus, as in the non-magnetic case, a D type ionization front
forms, gas moves radially outward, and a spherical dense shell
forms.  Ionization balance requires that when the shell bouding the
HII region reaches a radius $r_{\rm sh}$, the gas that remains in its
interior and is not in the shell wall must have started within a
distance
\begin{equation}
r_0 = \sqrt{r_s r_{\rm sh}} = r_s \left(1 +
\frac{7t}{4t_s}\right)^{2/7}
\end{equation}
of the ionizing source. Since the magnetic field is frozen into the
gas, the magnetic flux passing through the interior of the ionized
region is therefore
\begin{equation}
\Phi_i = \pi B_n r_0^2 = \pi B_n r_s^2 \left(1 +
\frac{7t}{4t_s}\right)^{4/7},
\end{equation}
where $B_n$ is the magnitude of the magnetic field in the neutral gas
outside the shell. In contrast, the total flux passing through the
ionized region and the dense shell combined is simply the initial flux
that passed through a circle of radius $r_{\rm sh}$,
\begin{equation}
\Phi_{\rm sh} = \pi B_n r_{\rm sh}^2 = \pi B_n r_s^2 \left(1 +
\frac{7t}{4t_s}\right)^{8/7}.
\end{equation}
Since the magnetic flux through the HII region interior rises as
$t^{4/7}$ at late times, while the total flux through the HII
region interior and the dense shell varies as
$t^{8/7}$, the fraction of magnetic field lines passing through the
HII region that pass through the shell interior must decline with time
as $t^{-4/7}$. Physically, this occurs because magnetic field lines
are being dragged with the gas out of the HII region interior and
concentrated in the dense shell.

As time progresses, the magnetic field changes the evolution in two
ways. First, magnetic pressure and tension oppose the expansion of the
shell perpendicular to the magnetic field, causing it to deform and
become aspherical. This effect should become significant when magnetic
pressure is comparable to thermal pressure in the ionized gas,
i.e. when $\rho_n v_A^2 \sim \rho_i c_i^2$. Since $\rho_i\approx
(r_{\rm sh}/r_s)^{-3/2} \rho_n$ before magnetic effects are
significant, we should see significant deformation of the HII region
when it reaches a radius of order $r_{\rm sh} \approx
(c_i/v_A)^{4/3} r_s$, and the shell velocity is $v_{\rm sh} \approx
v_A$. Second, magnetic pressure in the swept-up shell will limit
compression in the shock perpendicular to the magnetic field to a
factor of order $v_{\rm sh}/v_A$. This will make the shell thinner
along the field and thicker perpendicular to it, an effect that will
become noticable when $v_{\rm sh} \sim v_A$. Thus, we expect both
magnetic effects to become significant when the region has expanded to
a radius of roughly 
\begin{equation}
r_m \equiv \left(\frac{c_i}{v_A}\right)^{4/3} r_s,
\end{equation}
where we have defined the magnetic critical radius $r_m$ for the HII
region. Assuming $r_m \gg r_s$, this happens at roughly a time
\begin{equation}
t_m \equiv \frac{4}{7}\left(\frac{c_i}{v_A}\right)^{7/3} t_s
\end{equation}
after the start of the evolution. 

The evolution will then enter a new stage in which the thermal
pressure inside the ionized region is small compared to the magnetic
pressure, but both still greatly exceed the thermal pressure in the
neutral gas. Since $|\vecB|^2 \gg \rho_i c_i^2$, gas will be able to
move only along field lines, and those field lines will be
approximately straight. Since expansion is subsonic with respect to
the ionized gas and we are concerned with times $\gg t_s$, the density
in the ionized gas along each field line is constant. Along the field,
gas motions are unrestricted and the evolution will be a normal D type
ionization front as seen in non-magnetized gas. Perpendicular to the
field, magnetic pressure effects become more and more dominant as time
goes on. The ionized gas continues to
escape along field lines, so the pressure driving expansion
perpendicular to the field continues falling, which causes the front
to slow and the density contrast to decrease perpendicular to the
field. At the same time, everywhere that the magnetic field is at all
oblique to the front, a slow mode will move into the neutral
medium. This mode will remove magnetic support from the front by
bending the field lines so that they become perpendicular to the
shock \citep{williams00a}. This loss of magnetic support will cause
the shell to transition from thick and magnetic pressure-dominated to
thin and without much magnetic support over a larger and larger solid
angle as time goes on, until only the circle exactly perpendicular to
the initial magnetic field is not occupied by a dense shell.

\subsection{Simulation}

We simulate an initially neutral uniform medium with density $\rho_n =
2.34\times 10^{-22}$ g cm$^{-3}$ ($n_H = 100$ cm$^{-3}$) and
temperature $T_n = 11$ K, threaded by
a magnetic field $\vecB = 14.2 \,\hat{\mathbf{x}}$
$\mu$G ($4.0 \times 10^{-6}$ in the units we use in this paper). This
gives an Alfven speed of 2.6 km s$^{-1}$ and a sound speed of 0.20 km
s$^{-1}$ in the gas at the initial time. The computational domain runs
from $-10$ pc to $10$ pc in every direction, and there is an ionizing
source at the origin with luminosity $s=4.0\times 10^{46}$ s$^{-1}$,
giving an initial Str\"omgren radius $r_s = 0.5$ pc and sound crossing
time $t_s = 0.056$ Myr, assuming a sound speed $c_i = 8.7$ km s$^{-1}$
in the ionized region. For these parameters the magnetic critical
radius is $r_m = 6.0 r_s = 2.5$ pc and the magnetic critical time is
$t_m = 9.4 t_s = 0.53$ Myr. The simulation uses a resolution of
$256^3$ cells and a heating/cooling time step constraint of $f=4$. We
run the simulation for a time of $3 t_m = 1.58$ Myr.

\subsubsection{Structure at $t\ll t_m$}

\begin{figure}
\plotone{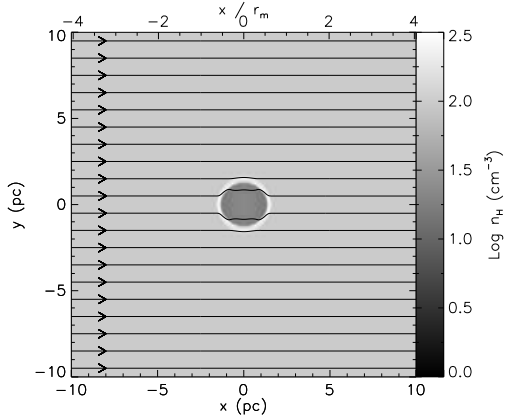}
\caption{
\label{magearly1}
Slice through the $xy$ plane showing log of hydrogen number
density in a simulation of an ionization front expanding into a
magnetized medium after a time of roughly $t_m/3=0.177$ Myr. The lines
are magnetic field lines. The density of lines outside of the HII
region corresponds to a magnetic field strength of $14.2$ $\mu$G.
}
\end{figure}

\begin{figure}
\plotone{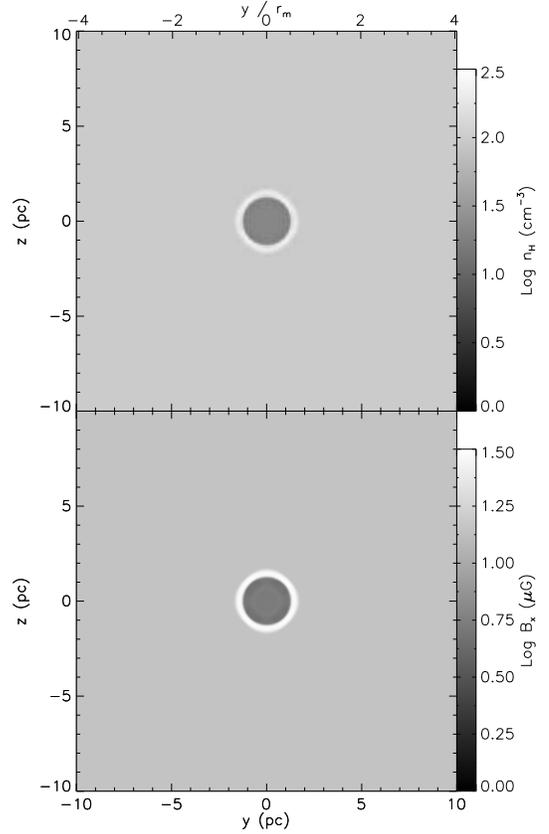}
\caption{
\label{magearly2}
Slices through the $yz$ plane showing log of hydrogen number
density (\textit{upper panel}) and log of magnetic field strength
(\textit{lower panel}) in a simulation of an ionization front
expanding into a magnetized medium after a time of roughly
$t_m/3=0.177$ Myr.
}
\end{figure}

Figures \ref{magearly1} and \ref{magearly2} show slices through the
computational domain at a time of $t\approx t_m/3=0.177$ Myr. As
expected, the HII
region almost spherical, although there is already a slight asymmetry
visible. The outer boundary of the dense shell is nearly perfectly
round, and is at essentially the same radius as we would expect for a
non-magnetized region at the same time. However, the shell is slightly
thicker in the direction perpendicular to the field than in the
direction along the field, due to magnetic pressure opposing
compression. Also as expected, because magnetic field lines are being
advected with the gas, the magnetic field inside the HII region is
much weaker than the background field, and the magnetic field in the
dense shell is much stronger than the background field.

\subsubsection{Structure at $t\sim t_m$}

\begin{figure}
\plotone{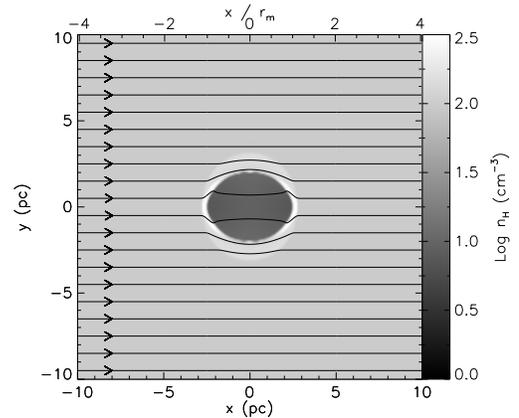}
\caption{
\label{magmid1}
Slice through the $xy$ plane showing log of hydrogen number
density in a simulation of an ionization front expanding into a
magnetized medium after a time of roughly $t_m=0.53$ Myr. The lines
are magnetic field lines. The density of lines outside of the HII region
corresponds to a magnetic field strength of $14.2$ $\mu$G.
}
\end{figure}

\begin{figure}
\plotone{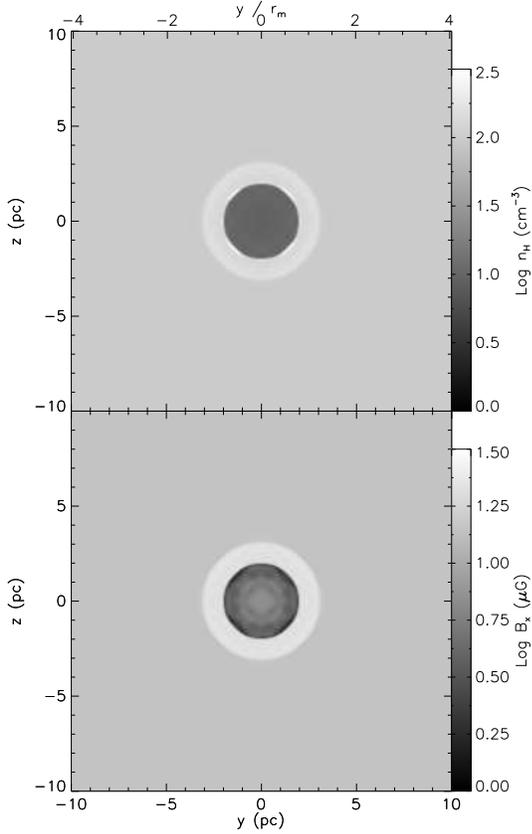}
\caption{
\label{magmid2}
Slices through the $yz$ plane showing log of hydrogen number
density (\textit{upper panel}) and log of magnetic field strength
(\textit{lower panel}) in a simulation of an ionization front
expanding into a magnetized medium after a time of roughly $t_m=0.53$
Myr.
}
\end{figure}

Figures \ref{magmid1} and \ref{magmid2} show the same slices as
Figures \ref{magearly1} and \ref{magearly2} at a time
$t\approx t_m= 0.53$ Myr. As expected, the magnetic effects have now become
pronounced. The inner, ionized region is strongly prolate, so that it
is roughly a factor of 2 longer along the field than perpendicular to
it. Along the field, the dense shell is still only a few cells
thick, the same as in the non-magnetic case, but in the directions
perpendicular to the field its thickness is not much smaller than its
radius, again as we expect. However, the outer edge of the expanding shell is
still not too far from spherical, and in the direction along the field its
radius is close to $r_m$, the radius we would expect were there no
field present. 

Interestingly, the shell extends
somewhat further in the direction across the field than in the
direction along it, so the outer radius of the swept up shell in the
$y$ and $z$ directions is noticably larger than $r_m$. 
This is likely because at $t=t_m$ the expansion
speed is comparable to the Alfven speed, and is slower than the fast
magnetosonic speed. Across the field, signals from the shell can
propogate by fast magnetosonic waves more rapidly than the shell can
expand along the field. This creates a density disturbance, carried by
the fast mode, ahead of the
radius that the shell would have reached had there been no magnetic
field. Since fast waves cannot propogate parallel to the magnetic
field, the leading edge of the disturbance has advanced less in the
$x$ direction than in the $y$ or $z$ directions.

\subsubsection{Structure at $t\gg t_m$}

\begin{figure}
\plotone{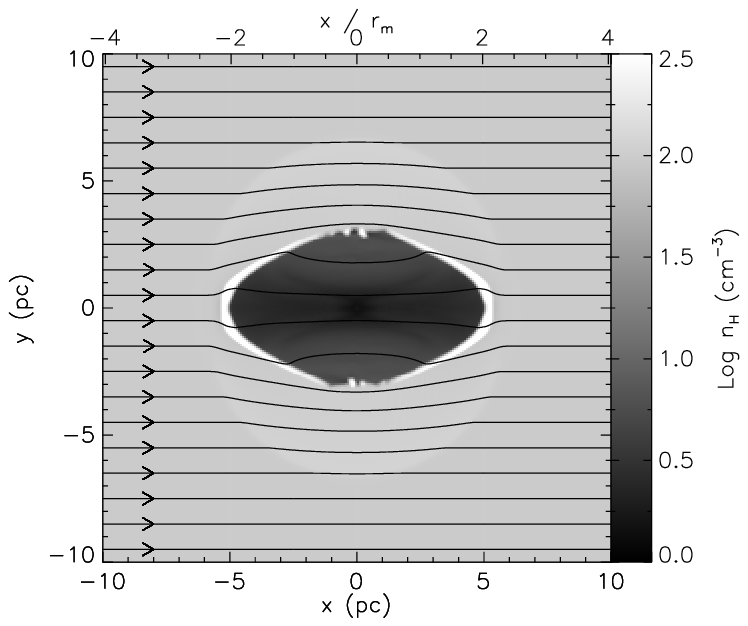}
\caption{
\label{maglate1}
Slice through the $xy$ plane showing log of hydrogen number
density in a simulation of an ionization front expanding into a
magnetized medium after $1.58$ Myr ($3 t_m$). The lines are magnetic
field lines. The density of lines outside of the HII region
corresponds to a magnetic field strength of $14.2$ $\mu$G. 
}
\end{figure}

\begin{figure}
\plotone{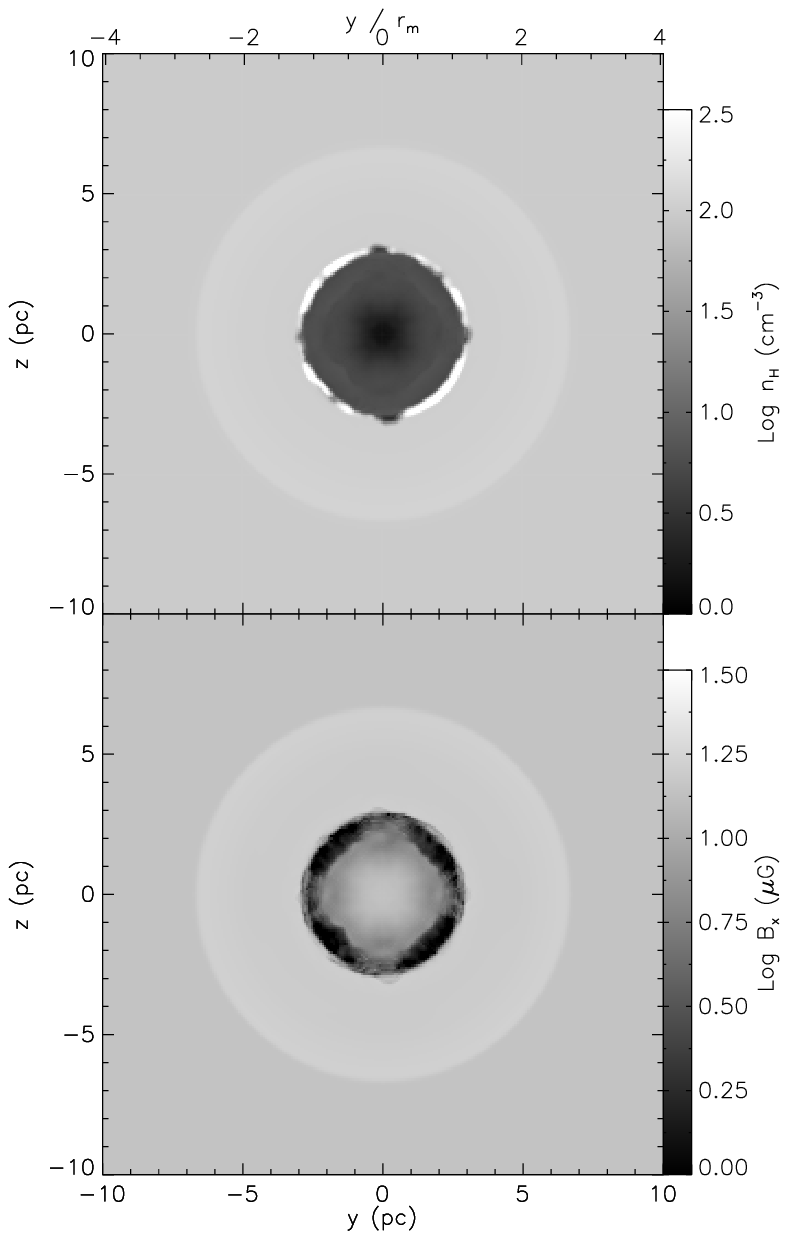}
\caption{
\label{maglate2}
Slices through the $yz$ plane showing log of hydrogen number
density (\textit{upper panel}) and log of magnetic field strength
(\textit{lower panel}) in a simulation of an ionization front
expanding into a magnetized medium after $1.58$ Myr ($3 t_m$). 
}
\end{figure}

Figures \ref{maglate1} and \ref{maglate2} show the simulation at our
final time slice, $t=1.58\mbox{ Myr} =3 t_m$. In the evolution after
time $t_m$, the inner ionized region becomes more prolate,
expanding more rapidly along the field than perpendicular to
it. Along the field, in the $x$ direction, there continues to be dense
shell of swept up material.

In the $y$ and $z$ directions, perpendicular to the field lines,
the thick shell of gas bounded between
the ionization front and the fast mode front, has also continued to
expand. However, its overdensity is decreasing, and at this time is
only $120$ cm$^{-3}$, so it is within $20\%$ of
the background density, and smaller than the density at time
$t_m$. The shell expansion velocity perpendicular to the field lines
is much smaller than the Alfven speed in the neutral gas, and is
comparable to the sound speed, so the leading edge of the shell is no
longer a shock. Instead, the shell is just a small overdensity that is
gradually relaxing away.

The boundary between the neutral and ionized regions in the $y$ and $z$ directions appears to be somewhat corrugated, but this may well be a numerical artifact, particularly since Figure \ref{maglate2} shows that the structure at the ionization front is aligned with the computational grid. It may be that for ionization fronts that are nearly static, as this one is due to magnetic confinement, rebuilding the ray tree once every five time steps is insufficient to prevent the growth of structures.

In between the $x$ direction and the $y$ direction, there is a region
where the field is oblique relative to the ionization front. As
predicted by \citet{williams00a}, this region shows two types of
behavior depending on how close the field is to being parallel or
perpendicular to the front. Closer to the $x$ direction, the field
been bent so it is close to perpendicular to the front, and a dense
shell bounded has formed at the ionization front. Closer to the $y$
direction, the field is not bent and enters the front at an oblique
angle, and there is no shock or dense shell. The solid angle over
which a dense shell exists appears to be increasing in time.

While expansion of the ionized region perpendicular to the field has
slowed, the expansion velocity along the field lines is actually
larger or about the same as it was at time $t_m$. To understand the
origin of this
effect, consider the jump conditions that govern the expansion of the
dense shell in the $x$ direction at late times, when the expansion is
strongly subsonic with respect to the ionized gas, but the thermal
pressure in the ionized gas is still much greater than the thermal
pressure in the neutral gas. In the rest frame of the shell, this
means that ram pressure of neutral gas flowing into the shell must
balance the thermal pressure of ionized gas downstream from it. Thus,
$\rho_n v_{\rm sh}^2 \propto \rho_i c_i^2$, so the shell expansion
velocity obeys $v_{\rm sh} \propto c_i \sqrt{\rho_i/\rho_n}$. Thus far
our analysis could apply just as well to a non-magnetized HII
region. The difference that the magnetic field makes is that it
greatly reduces the expansion rate of the ionized region in two
directions. This prevents the density from dropping as fast as in the
hydrodynamic case, which keeps the expansion velocity larger at later
times.

However, while the decline in ionized gas density and pressure is
slower than in the hydrodynamic case, since gas continues to flow out
of the ionzied region along field lines into the end caps, the
pressure inside the HII region
is always decreasing after time $t_m$. By $3\,t_m$, the HII region is
in the limit where the magnetic pressure inside the ionization
front is higher than the thermal pressure. As a result, the field
lines have begun to straighten out inside the ionized region, so they
are kinked only at the dense end caps. The magnetic energy density
inside the ionized region is actually larger than at time $t_m$, since
for $t\gtsim t_m$ field lines that were pushed into the dense shell
start to move back into the ionized region interior.

\section{Discussion and Conclusion}
\label{conclusion}

We have demonstrated an algorithm for computing the evolution of
magnetized molecular gases subjected to internal sources of ionizing
radiation, which is potentially applicable to molecular clouds. In
testing our algorithm, we have discovered three conditions that are
likely to apply to ionizing radiation hydrodynamic and
magnetohydrodynamic codes in general. First, to achieve maximum
accuracy the update time step must be limited so that the temperature
in cells does not change by more than a factor of $f<10$ between
hydrodynamic or magnetohydrodynamic updates. Larger values of $f$
produce small but significant errors in the expansion rates of
D type ionization fronts. Second, when
using a ray-tracing approach to compute the ionizing radiative
transfer, one should rotate the orientation of the rays periodically to
avoid a build-up of errors caused by the discretization of angles
around the ionizing source. Failure to obey this conditions results in
fronts that should be spherical developing aspherical features, and
in more complex calculations this could potentially seed
instabilities. Third, and most significantly, one must avoid
overcooling caused by numerical smearing of the ionization front. This
can be handled most easily by suppressing cooling in cells with mixed
ionization fractions. Failure to obey this condition leads to an
unphysical loss of energy from expanding HII regions that causes them
to lag analytic solutions by tens of percent. A calculation that
satisfies these three constraints can reproduce the analytic solution
for the expansion of a D type ionization front to an accuracy of a
percent for at least $\sim 20$ ionized sound-crossing times.

Using our algorithm, we report the first three-dimensional simulations
of the expansion of an HII region into a magnetized gas. We show that
the presence of a magnetic field distorts the HII region, and
greatly reduces the strength of the shock and the density contrast in
directions perpendicular
to the magnetic field. This leads to the formation of an HII region
which is bounded by a dense shell of swept-up gas along the field, but
not perpendicular to it. The absence of a dense shell over much of the
solid angle means that, in the presence of strong, ordered magnetic
fields, HII regions may not be able to collect and compress as much
gas as one might expect from purely hydrodynamic estimates. This may
reduce the efficiency of triggered star formation from HII
regions.

\acknowledgements 
We thank R. Crockett, R. Klein,  C. McKee,
E. Ostriker, and D. Whalen for helpful discussions, and the referee
for comments that improved the paper. Support for this
work was
provided by NASA through Hubble Fellowship grant \#HSF-HF-01186
awarded by the Space Telescope Science Institute, which is operated by
the Association of Universities for Research in Astronomy, Inc., for
NASA, under contract NAS 5-26555 (MRK) and by award DE-FG52-06NA26217
from the DOE (JMS). This research used
computational facilities supported by NSF grant AST-0216105. The
authors are pleased to acknowledge that the work reported on in this
paper was substantially performed at the TIGRESS high performance
computer center at Princeton University which is jointly supported by
the Princeton Institute for Computational
Science and Engineering and the Princeton University Office of
Information Technology.

%\bibliographystyle{apj}
%\bibliography{refs}

%--------------------------------------------------------------------------
\end{document}